\documentclass[a4paper,11pt,bbold,amssymb,amsmath]{article}
\pdfoutput=1 

\usepackage{jheppub}

\newcommand{\beqn}{\begin{eqnarray}}
\newcommand{\eeqn}{\end{eqnarray}}
\newcommand{\eq}[1]{(\ref{#1})}

\newcommand{\eff}{{B\Omega}}

\newcommand{\CMW}{{{\mbox{\tiny{CMW}}}}}
\newcommand{\CVW}{{\mbox{\tiny{CVW}}}}
\newcommand{\CHW}{{\mbox{\tiny{CHW}}}}
\newcommand{\VH}{{\mbox{\tiny{VH}}}}
\newcommand{\MH}{{\mbox{\tiny{MH}}}}
\newcommand{\MV}{{\mbox{\tiny{MV}}}}
\newcommand{\MVH}{{\mbox{\tiny{MVH}}}}
\newcommand{\DHS}{{\mbox{\tiny{DHS}}}}

\newcommand{\bs}{\boldsymbol}

\newcommand{\unity}{\bbbone}
\def\bbbone{{\mathchoice {\rm 1\mskip-4mu l} {\rm 1\mskip-4mu l} {\rm 1\mskip-4.5mu l} {\rm 1\mskip-5mu l}}}

\title{Chiral Heat Wave and mixing of Magnetic, Vortical and Heat waves in chiral media}

\author{M.~N.~Chernodub}
\affiliation{CNRS, Laboratoire de Math\'ematiques et Physique Th\'eorique, Universit\'e de Tours, 37200 France}
\affiliation{Soft Matter Physics Laboratory, Far Eastern Federal University, Sukhanova 8, Vladivostok, Russia}
\affiliation{Department of Physics and Astronomy, University of Gent, Krijgslaan 281, S9, Gent, Belgium}
\emailAdd{maxim.chernodub@lmpt.univ-tours.fr}

\abstract{We show that a hot rotating fluid of relativistic chiral fermions possesses a new gapless collective mode associated with coherent propagation of energy density and chiral density waves along the axis of rotation. This mode, which we call the Chiral Heat Wave, emerges due to a mixed gauge-gravitational anomaly. At finite density the Chiral Heat Wave couples to the Chiral Vortical Wave while in the presence of an external magnetic field it mixes with the Chiral Magnetic Wave. The coupling of the Chiral Magnetic and Chiral Vortical Waves is also demonstrated. We find that the coupled waves -- which are coherent fluctuations of the vector, axial and energy currents -- have generally different velocities compared to the velocities of the individual waves.}

\begin{document} 
\maketitle
\flushbottom

\section{Introduction}

Anomalies in chiral matter lead to unusual transport effects in an impressive variety of physical systems and energy scales. The Chiral Magnetic Effect (CME) generates a vector (electric) current along an external magnetic field in a chirally imbalanced matter~\cite{ref:CME:1,ref:Vilenkin:CME,Alekseev:1998ds}. The Chiral Separation Effect (CSE) implies the existence of an axial (chiral) current along the background magnetic field in dense chiral systems~\cite{ref:CSE:1,ref:CSE:2}. In a rotating fluid or plasma of chiral fermions the Chiral Vortical Effects (CVE) lead to the appearance of both vector and axial currents along the axis of rotation~\cite{ref:CVE:1,ref:CVE:2,ref:CVE:3,ref:CVE:4,ref:CVE:5}. Finally, both magnetic field and global rotation of the chiral matter produce an energy flux parallel to the corresponding axes~\cite{ref:Energy:1,ref:AME:lattice}. All these effects originate either from the chiral anomaly or from the mixed gauge-gravitational anomaly. 
 
The anomalous transport phenomena are expected to be realized in the quark-gluon plasma created in heavy-ion collisions~\cite{ref:CME:QGP:1,ref:CME:QGP:2,ref:CME:QGP:3}, in liquid Helium ${}^3$He-A~\cite{Volovik:1999wx,Volovik2003}, in Dirac~\cite{ref:CME:Dirac:1,ref:CME:Dirac:2,ref:CME:Dirac:3} and Weyl~\cite{ref:CME:Weyl,ref:AME:Weyl} semimetals, in cold atomic gases~\cite{ref:gases}, in Early Universe~\cite{ref:Vilenkin:Astro,Boyarsky:2011uy}, in neutron stars and supernovae~\cite{ref:stars:1,ref:stars:2,ref:stars:3}. Recent reviews on anomalous transport phenomena can be found in Ref.~\cite{ref:reviews:1,ref:reviews:2,ref:reviews:3,ref:reviews:4,ref:reviews:5}.

In a stationary uniform background the anomalies generate steady time-independent currents. However, the chiral matter is also known to support certain types of sound-like density waves associated with anomalous transport because the anomalous transport laws couple vector and axial charge densities and their currents to each other and force their perturbations to be interrelated. As a result, the vector and axial charge densities may propagate as a common vector-axial density wave.  

In an external magnetic field the anomalous coupling between the vector and axial sectors is given by the CME and CSE. The corresponding gapless excitation is called the Chiral Magnetic Wave (CMW)~\cite{Newman:2005hd,ref:CMW}. The CMW propagates along the axis of magnetic field with the velocity which depends on the magnetic field strength. This wave, which may exist even at zero fermion density, was suggested to reveal itself in heavy-ion collisions via the electric quadrupole observables~\cite{Gorbar:2011ya,Burnier:2011bf}. Certain experimental signatures consistent with possible existence of the CMW in quark-gluon plasma were indeed found recently~\cite{Adamczyk:2015eqo}, although there are arguments suggesting that these signatures may have another explanation~\cite{Hatta:2015hca}.

If a finite-density chiral system is set into rotation then the vector and axial charge densities are coupled to each other by the CVE. The associated gapless mode is called the Chiral Vortical Wave (CVW)~\cite{ref:CVW}. The CVW propagates along the axis of rotation and it may in principle be observed in noncentral heavy ion collisions which create rotating quark-gluon plasmas. 

Another gapless excitation, the Chiral Alfv\'en Wave~\cite{ref:CAW}, corresponds to sound-like oscillations of local velocity of a charged chiral fluid in the presence of an external magnetic field. Despite the fluid oscillations are transverse with respect to the magnetic field axis, the wave itself propagates along magnetic field lines in a close analogy with the usual Alfv\'en modes that exist in various ion plasmas. 

An external electric field may also lead to appearance of a gapless mode due to the so-called Chiral Electric Separation Effect (CESE). The CESE generates the axial current in a chirally imbalanced medium in the presence of an external electric field. The associated sound mode should propagate along the axis of the electric field and may be potentially observable in heavy-ion collisions~\cite{Huang:2013iia}.

In our paper we demonstrate the existence of a new gapless collective mode associated with a coherent propagation of the (thermal) energy wave and the (chiral) axial density wave in a globally rotating medium of relativistic chiral fermions. We call this collective mode the Chiral Heat Wave (CHW). 

The heat wave is substantially different from the magnetic and vortical waves\footnote{We discuss three chiral waves with visually similar abbreviations (CMW, CVW, CHW) as well as various coupled (CMW and/or CVW and/or CHW) waves. Thus, in order to avoid almost inevitable confusion, we often use the terms ``magnetic wave'', ``vortical wave'', ``heat wave'', ``magnetic-vortical wave'' etc.}.  For example, the heat wave may be realized in the absence of a magnetic field background unlike the magnetic wave.  Moreover, the heat wave may propagate in a zero density system unlike the chiral vortical wave. Finally, as we mentioned, the pure Chiral Heat Wave couples thermal energy waves with axial density waves, while the Chiral Magnetic and and Chiral Vortical Waves -- in their original formulation -- correspond to a coherent propagation of vector and axial density waves in the absence of thermal energy waves.

The structure of this paper is as follows. In Sect.~\ref{sec:CMW} we review in detail both the Chiral Magnetic and the Chiral Vortical Waves. In Sect.~\ref{sec:CHW} we introduce the Chiral Heat Wave and discuss its basic properties. As we show in Sect.~\ref{sec:mix} these waves may mix in different combinations and propagate as a common vector-axial-energy density wave. The properties (velocity, direction, density content) depend on the concrete physical environment (density, temperature, global rotation and magnetic field). A special case of zero-temperature energy waves (basically, matter waves) is also discussed. In Sect.~\ref{sec:DHS} we describe a special class of non-propagating diffusion modes which we call the Dense Hot Spots (DHS). These long-wavelength configurations carry nonzero vector charge and energy density while their axial charge is zero. They may only appear in rotating systems subjected to an external magnetic field. Finally, Sect.~\ref{sec:summary} is devoted to a summary of our results and discussion.

\section{Chiral Magnetic Wave and Chiral Vortical Wave}
\label{sec:CMW}

\subsection{Chiral Magnetic Wave}

In this Section we briefly review the Chiral Magnetic Wave (CMW) following Ref.~\cite{ref:CMW}. The CMW is a collective gapless excitation in a system (fluid) of massless charged fermions in the background of external magnetic field ${\bs B}$. The CMW appears due to correlated interplay of the CME and the CSE, which describe, respectively, the dissipationless transfer of electric charge and chiral charge along the magnetic field~\cite{ref:CME:1,ref:CSE:1,ref:CSE:2}:
\beqn
{\bs j}_V & = & \sigma_{V}^{{\cal B}} e {\bs B} \,, 
\label{eq:CME:jV}\\
{\bs j}_A & = & \sigma_{A}^{{\cal B}} e {\bs B}\,,
\label{eq:CSE:jA}
\eeqn
The CME~\eq{eq:CME:jV} generates electric (vector) current of the fermions, ${\bs j}_V \equiv {\bs j}$ along the direction of magnetic field, while the CSE \eq{eq:CSE:jA} leads to appearance of the chiral (axial) current ${\bs j}_A \equiv {\bs j}_5$ given, respectively, by a sum and a difference of the right-handed (${\bs j}_R$) and left-handed (${\bs j}_L$) fermionic currents:
\beqn
\begin{split}
{\bs j}_V  & = {\bs j}_R + {\bs j}_L\,, \\ 
{\bs j}_A  & = {\bs j}_R - {\bs j}_L\,.
\end{split}
\label{eq:currents}
\eeqn

The strength of these effects is controlled by the corresponding anomalous transport coefficients in Eqs.~\eq{eq:CME:jV} and \eq{eq:CSE:jA}:
\beqn
\begin{split}
\sigma_{V}^{{\cal B}} & = \frac{\mu_A}{2 \pi^2}\,, \\ 
\sigma_{A}^{{\cal B}} & = \frac{\mu_V}{2 \pi^2}\,, 
\end{split}
\label{eq:sigma:VB:AB}
\eeqn
where $\mu_V$ is the usual (vector) chemical potential which describes the total density of the right-handed and left-handed
fermions while $\mu_A $ is the axial (chiral) chemical potential which describes the difference in their densities, respectively:
\beqn
\begin{split}
\mu_V \equiv \mu & = \frac{1}{2}(\mu_R + \mu_L)\,, \\
\mu_A \equiv \mu_5 & = \frac{1}{2}(\mu_R - \mu_L)\,.
\end{split}
\eeqn
The superscripts ${\cal B}$ in Eq.~\eq{eq:sigma:VB:AB} indicate that these transport coefficients correspond to the background magnetic field~${\bs B}$.

Let us consider small long-wave perturbations in the vector and axial charge densities, $\rho_V \equiv j^0_V$ and $\rho_A \equiv j^0_A$, respectively. These perturbations,
\beqn
\delta \rho_V(x) & = & \chi_{VV} \delta \mu_V(x) + \chi_{VA} \delta \mu_A(x)\,, 
\label{eq:delta:qV}\\
\delta \rho_A(x) & = & \chi_{AV} \delta \mu_V(x) + \chi_{AA} \delta \mu_A(x)\,,
\label{eq:delta:qA}
\eeqn
are related to the (local) deviations in the corresponding chemical potentials $\delta \mu_a = \mu_a - {\bar \mu}_a$ via the susceptibilities $\chi_{ab}$ with $a,b=V,A$. 
Here the bar over a quantity indicates a volume mean of the corresponding quantity\footnote{In order to avoid cluttering of our notations, hereafter we omit the bars over most of the mean quantities.}. 

Equations~\eq{eq:CME:jV}--\eq{eq:delta:qV} should be supplemented with the conservation laws for the vector and axial charges,
\beqn
\partial_\mu j^\mu_A & \equiv & \partial_t \rho_V + {\bs \partial} {\bs j}_V = 0\,,
\label{eq:cons:jV}\\
\partial_\mu j^\mu_A & \equiv & \partial_t \rho_A + {\bs \partial} {\bs j}_A = 0\,,
\label{eq:cons:jA}
\eeqn
where $j^\mu_V = (\rho_V,{\bs j}_V)$ and $j^\mu_A = (\rho_A,{\bs j}_A)$. Notice that, in general, the axial current $j^\mu_A$ is not conserved due to the chiral anomaly. However, the conservation law~\eq{eq:cons:jA} is valid in the absence of the external electric field which is the case considered in this paper.

For a totally neutral system of fermions, all chemical potential vanish on average: ${\bar \mu}_V = 0$ and ${\bar \mu}_A = 0$. Then we notice that $\rho_V$ and $\mu_V$ ($\rho_A$ and $\mu_A$) are the components of true (axial) vectors and therefore the covariance of Eqs.~\eq{eq:delta:qV} and \eq{eq:delta:qA} under the $P$ parity transformation ($V \to -V$ and $A \to A$) implies the absence of the off-diagonal terms in the susceptibility matrix, $\chi_{AV} = \chi_{VA} = 0$. Since we consider a chirally unbroken phase of the system, then the diagonal terms should be the same: $\chi_{VV} = \chi_{AA} = \chi$, where the susceptibility $\chi = \chi(|{\bs B}|)$ is a function of the magnetic field~\cite{ref:CMW}. This statement should also be true for ${ \mu}_V \neq 0$ provided that the chiral symmetry is unbroken. Therefore, for the chirally symmetric system (with $\mu_A = 0$) Eqs.~\eq{eq:delta:qV} and \eq{eq:delta:qA} are simplified:
\beqn
\begin{split}
\delta \rho_V(x) = \chi \delta \mu_V(x)\,,\\
\delta \rho_A(x) = \chi \delta \mu_A(x)\,.
\end{split}
\label{eq:delta:q:a}
\eeqn

From Eqs.~\eq{eq:CME:jV}, \eq{eq:CSE:jA} and \eq{eq:sigma:VB:AB} we deduce that a perturbation in electric and chiral charges leads to appearance of, respectively, perturbations in chiral and electric currents along the magnetic field: 
\beqn
\delta {\bs j}_V(x) & = & \frac{e{\bs B}}{2 \pi^2 \chi} \delta \rho_A(x)\,, 
\label{eq:CMW:jVz}\\[1mm]
\delta {\bs j}_A(x) & = & \frac{e {\bs B}}{2 \pi^2 \chi} \delta \rho_V(x)\,,
\label{eq:CMW:jAz}
\eeqn
while the transverse (with respect to the magnetic field) components of these currents are zero.

Let us now consider the uniform and constant magnetic field directed along the $z$ axis, ${\bs B} = B {\bs e}_z$. Without loss of generality one can take $eB>0$. Differentiating the currents~\eq{eq:CMW:jVz} and \eq{eq:CMW:jAz} over $z$ and applying the corresponding conservation laws~\eq{eq:cons:jV} and \eq{eq:cons:jA},
\beqn
\partial_t \delta \rho_V + \partial_z \delta j^z_V = 0\,, \qquad\quad
\partial_t \delta \rho_A + \partial_z \delta j^z_A = 0\,,
\label{eq:conservation:jVjA:z}
\eeqn
one gets the following system of linear differential equations which relates the perturbations in electric and chiral charge densities:
\beqn
\partial_t \delta \rho_V(x) + \frac{eB}{2 \pi^2 \chi} \partial_z \delta \rho_A(x) & = & 0\,, 
\label{eq:qVA}\\ 
\partial_t \delta \rho_A(x) + \frac{eB}{2 \pi^2 \chi} \partial_z \delta \rho_V(x) & = & 0\,.
\label{eq:qAV}
\eeqn
This system can easily be diagonalized by differentiating the first (second) equation with respect to $z$ or $t$ (with respect to $t$ or $z$, respectively) and combining them together. One obtains two gapless (massless) wave equations,
\beqn
\begin{split}
\left(\partial_t^2- v^2_{\CMW} \partial_z^2 \right) \delta \rho_V(x) = 0\,, \\[1mm]
\left(\partial_t^2- v^2_{\CMW} \partial_z^2 \right) \delta \rho_A(x) = 0\,, 
\end{split}
\eeqn
which describe a sound-like propagation of perturbations in electric and chiral charges (and their currents) along the axis of magnetic field with the velocity
\beqn
v_{\CMW} = \frac{eB}{2 \pi^2 \chi}\,.
\label{eq:v:CMW}
\eeqn
This gapless excitation is the Chiral Magnetic Wave (CMW).

The CMW is a coupled vector-axial density wave which propagates, qualitatively, as follows~\cite{ref:CMW}: 
\begin{itemize}

\item[(i)] due to the Chiral Magnetic Effect~\eq{eq:CMW:jVz} a local fluctuation of a chiral (axial) charge density generates a fluctuation in vector (electric) current in the direction of the magnetic field; 

\item[(ii)] in turn, the fluctuation in electric current leads, due to the electric charge conservation~\eq{eq:conservation:jVjA:z}, to a fluctuation of the electric charge density next point along the magnetic-field axis; 

\item[(iii)] next, the Chiral Separation Effect~\eq{eq:CMW:jAz} implies that the electric charge density generates the axial current which is again directed along the magnetic field axis; 

\item[(iv)] finally, due to the conservation of the axial charge~\eq{eq:conservation:jVjA:z} the fluctuation in the chiral current creates a fluctuation in the chiral charge density next point along the direction of magnetic field and then the whole processes repeats itself. 
\end{itemize}
Thus the Chiral Magnetic Wave is a chain-like process which involves the vector (electric) and axial (chiral) densities and their currents. 

Summing (subtracting) Eq.~\eq{eq:qAV} with (from) Eq.~\eq{eq:qVA} and using Eq.~\eq{eq:currents} we get two equations which describe propagation of fluctuations with definite chirality:
\beqn
\left(\partial_t \mp v_\CMW \partial_z \right) \delta \rho_{L,R} = 0\,,
\eeqn
where the upper (lower) sign corresponds to the left- (right-) handed fermions. Therefore the fluctuations in the densities of the left-handed (right-handed) fermions propagate opposite to (along) the direction of the magnetic field vector $e {\bs B}$. 

\begin{figure}[!thb]
\begin{center}
\includegraphics[scale=0.6,clip=true]{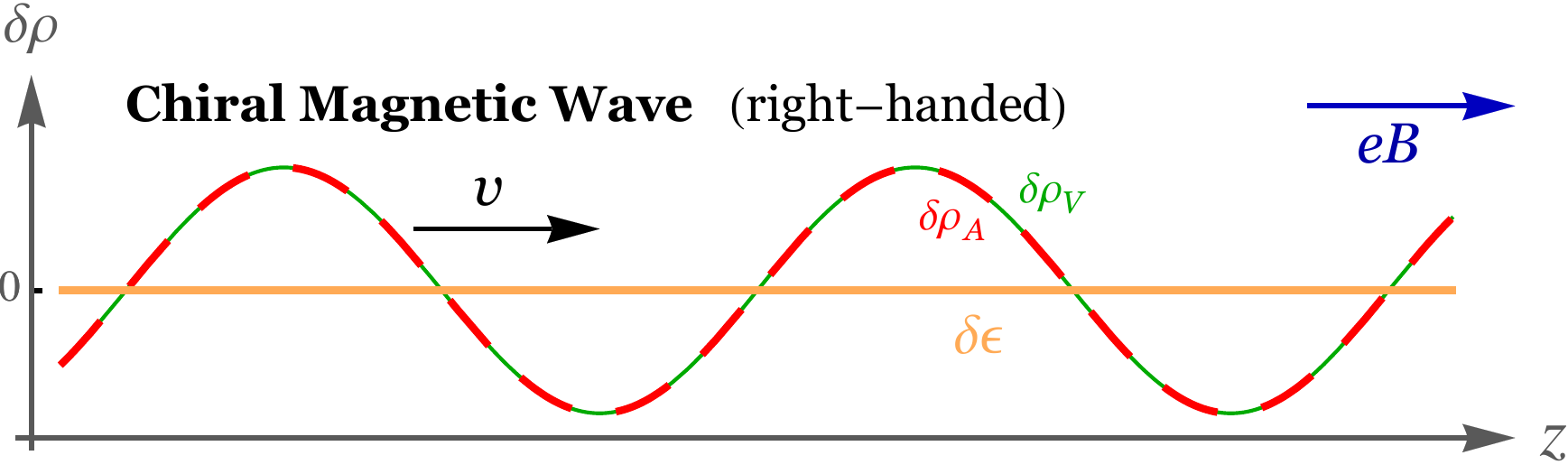}\\[5mm]
\includegraphics[scale=0.6,clip=true]{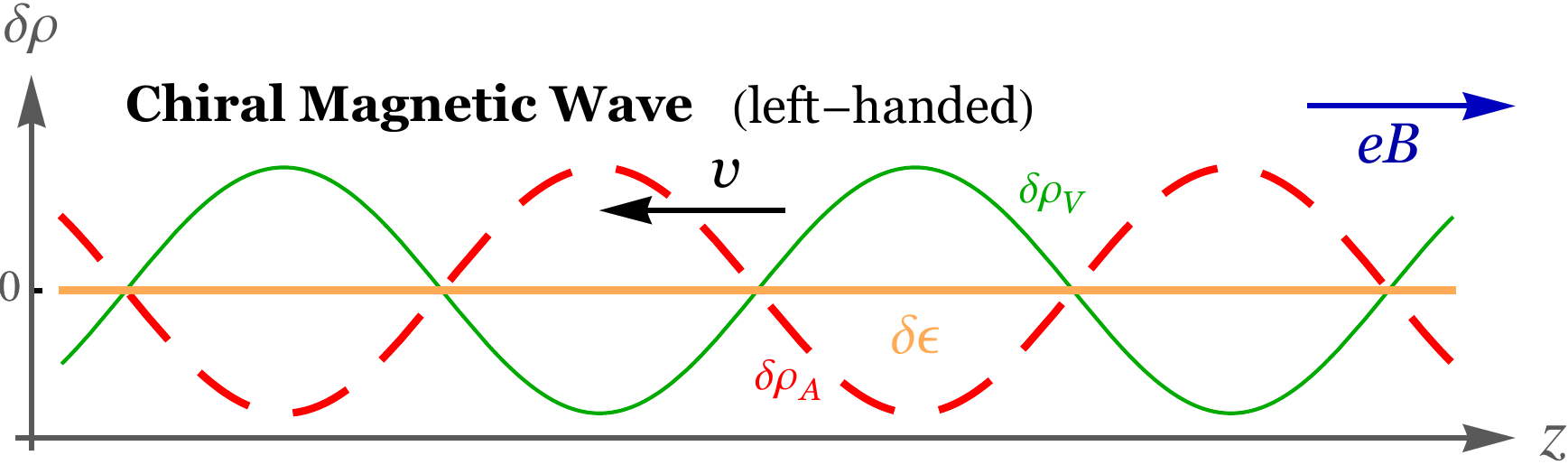}
\end{center}
\caption{Structure of the right-handed (the upper plot) and the left-handed (the lower plot) Chiral Magnetic Waves in terms of the vector (the green thin solid line), axial (the red thick dashed line) and energy (the thick solid orange line) densities for a zero-density ($\mu_V = 0$) cold ($T = 0$) nonrotating (${\bs \Omega} = 0$) chiral medium in the magnetic field background ${\bs B} \neq 0$. The vector ${\bs v}$ indicates the direction of the wave propagation.}
\label{fig:CMW}
\end{figure}

The structure of the CMW in terms of the individual density waves is schematically illustrated in Fig.~\ref{fig:CMW}. The energy component of the CMW will be discussed in Section~\ref{sec:MH}.

Before proceeding further we would like to make two important remarks, which concern both the Chiral Magnetic Wave discussed in this Section and other chiral waves mentioned below.

First, we notice that similarly to the CME and the CSE, the CMW is emerging at any strength of the magnetic field ${\bs B}$ since in the above derivation no assumption about the value of ${\bs B}$ was made~\cite{ref:CMW}. However, unlike the CME and the CSE, the CMW is a dissipative phenomenon. In the studied long-wave limit the dissipative terms should reveal themselves in the dispersion relations in the quadratic order in the longitudinal momentum~$k_z$:
\beqn
\omega_{L,R}(k_z) = \mp v_{\CMW} k_z - i D_\| k^2_z + \dots\,,
\label{eq:omega:LR}
\eeqn
where the first term describes the propagation of the CMW with velocity~\eq{eq:v:CMW} while the second term is responsible for its dissipation with a longitudinal diffusion constant $D_\|$. In Eq.~\eq{eq:omega:LR} the ellipsis stand for higher-order longitudinal terms and for transverse terms. Dissipative properties of the CMW were discussed in details in effective hydrodynamics and in holographic approaches~\cite{Jimenez-Alba:2014iia} and in chiral kinetic theory~\cite{Stephanov:2014dma}.

Second, in our paper we consider the Chiral Magnetic Wave and other waves in the absence of background flow of the chiral fluid. It is important to stress, however, that the fluid velocity may contribute to vector, axial and energy currents and affect the results of the paper if they are applied to realistic situations similar to the ones realized in quark-gluon plasmas. Realistic implementation of the Chiral Magnetic wave in heavy-ion collisions was considered in Ref.~\cite{Yee:2013cya}.

\subsection{Chiral Vortical Wave}

In this Section we briefly review the Chiral Vortical Wave (CVW) following Ref.~\cite{ref:CVW}. We consider a rotating fluid of massless fermions in the absence of magnetic field. A global rotation of the fluid can be expressed in terms of the vorticity ${\bs \Omega} = \frac{1}{2} {\bs \partial} \times {\bs v}$, where ${\bs v}$ is the local velocity of the fluid flow. Similarly to the CME and the CSE, the rotation should generate the vector and axial currents of the fermions along the axis of rotation~\cite{ref:CVE:1,ref:CVE:2,ref:CVE:3,ref:CVE:4,ref:CVE:5}:
\beqn
{\bs j}_V = \sigma_{V}^{{\cal V}} {\bs \Omega} \,, 
\qquad 
{\bs j}_A = \sigma_{A}^{{\cal V}} {\bs \Omega}\,,
\label{eq:CVE:jVA}
\eeqn
where the associated transport coefficients are a follows:
\beqn
\sigma_{V}^{{\cal V}} = \frac{\mu_V \mu_A}{\pi^2} \,, \qquad
\sigma_{A}^{{\cal V}} = \frac{T^2}{6} + \frac{\mu_V^2 + \mu_A^2}{2 \pi^2}\,,
\label{eq:sigma:CVE}
\eeqn
and $T$ is the temperature of the fluid. Equations~\eq{eq:CVE:jVA}--\eq{eq:sigma:CVE} describe the Chiral Vortical Effects (CVEs) in the first order of the angular velocity ${\bs \Omega}$. Throughout this paper we assume that the chiral fluid of relativistic fermions rotates slowly so that the linear approximation~\eq{eq:CVE:jVA} is valid. 

According to Eqs.~\eq{eq:CVE:jVA} and \eq{eq:sigma:CVE}, at finite temperature the rotating neutral fluid generates the axial current along the rotation axis ${\bs j}_A = T^2 {\bs \Omega}/6$. However,  unlike the transport coefficients for the CME and the CSE~\eq{eq:sigma:VB:AB}, the CVE coefficients depend quadratically on the chemical potentials and therefore in the neutral fluid an analogue of the CMW cannot appear.

However, if at least one of the chemical potentials is nonzero, then Eqs.~\eq{eq:CVE:jVA} and \eq{eq:sigma:CVE} indicate that fluctuation(s) $\delta \mu_{A,V}$ on top of the corresponding mean value(s) ${ \mu}_{A,V} \neq 0$ would couple linearly to the current(s) and may potentially lead to a wavelike excitation in a manner of the CMW that was discussed in the previous section. However, $\mu_A$ cannot be have a nonzero value in a realistic system in thermal equilibrium. Indeed, due to the existence of topological chirality--flipping processes the assumption of a nonzero value of the chiral density in thermodynamic equilibrium is not physical. Therefore, in our paper we always consider a system of chiral fermions with zero average chiral density (${\mu}_{A} = 0$). The vector density may, however, be nonzero (${ \mu}_{V} \neq 0$).
 
The CVE in a finite-density chirally-neutral rotating fluid generates a steady axial current,
\beqn
{\bar {\bs j}}_A = \left(\frac{T^2}{6} + \frac{{ \mu}_V^2}{2 \pi^2}\right) {\bs \Omega}\,.
\eeqn
while the average vector current vanish, ${\bar {\bs j}}_V = 0$. Expanding the CVE transport coefficients~\eq{eq:sigma:CVE} to the linear order in the fluctuations of the chemical potentials $\delta \mu_{V}$ and  $\delta \mu_{A}$ at fixed temperature $T$, one gets the following expression for the current fluctuations~\eq{eq:CVE:jVA}:
\beqn
\begin{split}
\delta {\bs j}_V & = \frac{{ \mu}_V}{\pi^2} \delta \mu_A\, {\bs \Omega} \equiv \frac{{ \mu}_V {\bs \Omega}}{\pi^2 \chi} \delta \rho_A\,  \,, \\
\delta {\bs j}_A & = \frac{{ \mu}_V}{\pi^2} \delta \mu_V \, {\bs \Omega}\equiv \frac{{ \mu}_V {\bs \Omega}}{\pi^2 \chi} \delta \rho_V\,  \,.
\end{split}
\label{eq:CVE:djVA}
\eeqn
Here we have used Eqs.~\eq{eq:delta:qV} and \eq{eq:delta:qA} assuming -- following the line of arguments for case of the CMW -- that the parity is unbroken in the fluid. In the above equations the susceptibility is, in general, a function $\chi = \chi({ \mu}_V, |{\bs \Omega}|)$ of the chemical potential ${ \mu}_V$ and the angular frequency~$\Omega$.

Equations~\eq{eq:CVE:djVA} demonstrate that in a rotating finite-density fluid, a fluctuation in the axial (vector) charge couples to the vector (axial) current exactly in the same manner as it happens in Eqs.~\eq{eq:CMW:jVz} and \eq{eq:CMW:jAz} which describe the CMW. Thus, the rotating fluid should also support a gapless wave-like excitation similar to the CMW. This excitation indeed exists and it is called the Chiral Vortical Wave (CVW)~\cite{ref:CVW}. 

The CVW has the same basic features as the CMW. It propagates along the axis of rotation ${\bs \Omega}$ with the velocity\footnote{Our definition~\eq{eq:delta:q:a} of the susceptibility $\chi$ differs from the one of the susceptibility $\chi_{\mu_0}$ of Ref.~\cite{ref:CVW} by the factor of two ($\chi \equiv 2 \chi_{\mu_0}$) since the relations $\delta \rho_{V/A} = \chi \delta \mu_{V/A}$ imply $\delta \rho_{L/R} = (\chi/2) \delta \mu_{L/R} \equiv  \chi_{\mu_0} \delta \mu_{L/R}$.}.
\beqn
v_{\CVW} = \frac{{ \mu}_V \Omega}{\pi^2 \chi}\,,
\label{eq:v:CVW}
\eeqn
which can easily be derived from Eq.~\eq{eq:v:CMW} by noticing that Eqs.~\eq{eq:CVE:djVA} differ from Eqs.~\eq{eq:CMW:jVz} and \eq{eq:CMW:jAz} by the simple substitution $e {\bs B} \to 2 { \mu}_V {\bs \Omega}$. Similarly to the magnetic wave, the right-handed (left-handed) chiralities in the vortical wave propagate along (opposite to) to the direction of the vorticity vector~$\mu_V {\bs \Omega}$.

\begin{figure}[!thb]
\begin{center}
\includegraphics[scale=0.6,clip=true]{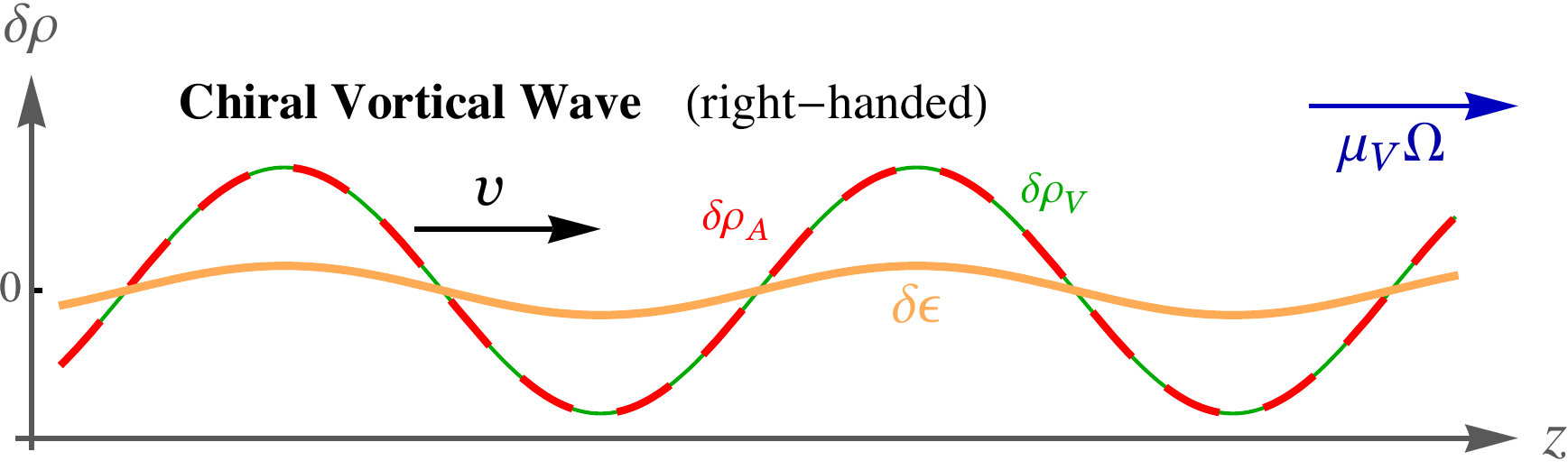}\\[5mm]
\includegraphics[scale=0.6,clip=true]{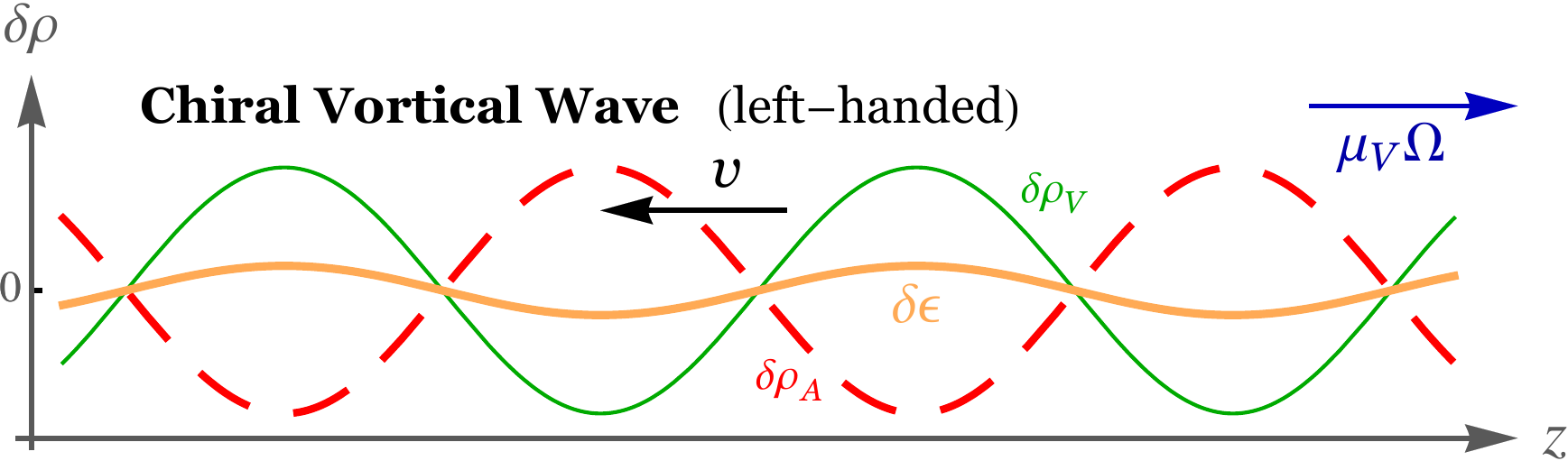}
\end{center}
\caption{Structure of the Chiral Vortical Wave in a dense ($\mu_V > 0$) cold ($T=0$) rotating (${\bs \Omega} \neq 0$) chiral medium at vanishing magnetic field (${\bs B} = 0$). If $\mu_V < 0$ then the energy density and the vector charge density have mutually opposite signs while the relative sign of the axial and vector densities remains the same. The notations are the same as in Fig.~\ref{fig:CMW}.}
\label{fig:CVW}
\end{figure}

The structure of the CVW in terms of the individual vector, axial and energy density waves is schematically shown in Fig.~\ref{fig:CVW}. The energy density content of the CVW will be discussed in Section~\ref{sec:VH}. At zero temperature the energy wave in the CVW is, basically, the mass wave~\eq{eq:epsilon:qV:mass:V} propagating in a uniform finite-density background given by $\mu_V \neq 0$.

 The CVW is suggested to exist in the quark-gluon plasma and its signatures are expected to be found in heavy-ion collisions~\cite{ref:CVW}.

\section{Chiral Heat Wave}
\label{sec:CHW}

\subsection{Nondissipative energy transfer}

The energy may also be transferred in a nondissipative way due to a mixed gauge-gravitational anomaly. The energy flux of a rotating fluid of massless fermions in a magnetic-field background is~\cite{ref:CVE:5,ref:Energy:1,ref:reviews:3}:
\beqn
{\bs j}_E & = & \sigma_{E}^{{\cal B}} e {\bs B} + \sigma_{E}^{{\cal V}} {\bs \Omega}\,,
\label{eq:jE}
\eeqn
where the energy current is given by components of the energy momentum tensor $T^{\mu\nu}$,
\beqn
j^i_E \equiv T^{0i} = \frac{i}{2} {\bar \psi} \left(\gamma^0 \partial^i + \gamma^i \partial^0\right) \psi\,,
\eeqn
and the anomalous conductivities are expressed via both chemical potentials and temperature:
\beqn
\sigma_{E}^{{\cal B}} & = & \frac{1}{2\pi^2} \mu_V \mu_A\,, 
\label{eq:sigma:EB}\\[1mm]
\sigma_{E}^{{\cal V}} & = & \frac{\mu_A}{3} \left[\frac{1}{\pi^2} \left( 3 \mu_V^2 + \mu_A^2 \right) + T^2 \right]\,.
\label{eq:sigma:EV}
\eeqn
The nondissipative energy transfer may only take place in the presence of a chiral imbalance (i.e., with $\mu_A \neq 0$) because the energy current ${\bs j}_E$ is a vector while both vorticity ${\bs \Omega}$ and the magnetic field ${\bs B}$ are pseudovectors. Thus, the energy current should only be related to ${\bs \Omega}$ and ${\bs B}$ by a coefficient which is linear in the pseudoscalar chemical potential $\mu_A$.

The energy conservation implies
\beqn
\partial_\mu T^{0\mu} \equiv \partial_t \epsilon + {\bs \partial} {\bs j}_E = 0\,,
\label{eq:energy:conservation}
\eeqn
where 
\beqn
\epsilon \equiv T^{00}\,,
\label{eq:epsilon}
\eeqn
is the (thermal) energy density. 

\subsection{Pure Chiral Heat Wave}

The aim of this paper is to find a new gapless wave-like excitation related to the anomalous energy transfer. In our context the energy waves should necessarily be related to local temperature fluctuations. Let us consider first a small deviation of temperature $\delta T$ (with $\delta T \ll {T}$) from its equilibrium value~${ T}$. In a linear approximation  the energy fluctuation $\delta \epsilon = \epsilon({ T}+ \delta T) - \epsilon({ T})$ is
\beqn
\delta \epsilon = c_V \delta T\,,
\label{eq:delta:epsilon}
\eeqn
where  $c_V \equiv c_V({ T})$ is the specific heat capacity:
\beqn
c_V = \left( \frac{\partial E}{\partial T}\right)_V\,.
\eeqn

In order to determine the nature of this wave, we notice that -- similarly to the cases of the CMW and the CVW --  the chiral chemical potential in equilibrium is zero, ${\bar \mu}_A = 0$, so that a nonzero value of $\mu_A$ may be due to fluctuations only. Let us also assume for a moment that the usual chemical potential is also zero, ${\bar \mu}_V = 0$ similarly to the simplest case of the  CMW. Then we find that in the linear approximation the coupling of the dissipationless energy current to the magnetic field~\eq{eq:sigma:EB} is quadratic in fluctuations so that the vector current ${\bs j}_A$ does not enter a linear wave equation that we search for. However, the coupling of the energy flow to the vorticity~\eq{eq:sigma:EV} has a linear term, $\sigma_{E}^{{\cal V}} = T^2 \delta \mu_A/3$, which is sensitive to the axial charge fluctuations. Thus, in a finite-temperature rotating fluid the energy fluctuations should couple to the chiral charge fluctuations. 

The fluctuations of the chiral charge may also couple to electric charge fluctuations in rotating fluid at finite magnetic field via the CMW~\cite{ref:CMW} and in rotating fluid at nonzero chemical potential via CVW~\cite{ref:CVW}. In order to demonstrate the existence of the new, energy-chiral charge wave, we consider first the rotating (${\bs \Omega} \neq 0$) finite-temperature ($T \neq 0$) system in the absence of magnetic field, ${\bs B} = 0$ (thus the CMW does not exist) and at zero density, $\mu_V = 0$ (thus the CVW does not exist either). In this environment only a pure energy-chiral charge wave may propagate as a collective excitation.

According to the full system of equations \eq{eq:CVE:jVA}, \eq{eq:sigma:CVE} and  \eq{eq:jE}, the fluctuations obey the following relations:
\beqn
\delta {\bs j}_V & = & 0\,, 
\label{eq:mixing1:jV:2}\\
\delta {\bs j}_A & = & \frac{{ T} {\bs \Omega}}{3} \, \delta T, 
\label{eq:mixing2:jA:2}\\
\delta {\bs j}_E & = & \frac{{ T}^2 {\bs \Omega}}{3} \delta \mu_A \,.
\label{eq:jE:2}
\eeqn
We notice first that the new wave does not have the vector density component~\eq{eq:mixing1:jV:2}. Then, using Eqs.~\eq{eq:delta:q:a} and \eq{eq:delta:epsilon} we rewrite the system~\eq{eq:mixing2:jA:2} and \eq{eq:jE:2} in the following form:
\beqn
\begin{split}
\delta {\bs j}_A & = \frac{{ T} {\bs \Omega}}{3 c_V} \, \delta \epsilon, \\
\delta {\bs j}_E & = \frac{{ T}^2 {\bs \Omega}}{3 \chi} \delta \rho_A \,,
\end{split}
\label{eq:mixing2:jAE:3}
\eeqn
which clearly demonstrates the coupling of the chiral (energy) current to the energy (chiral) density perturbations, similarly to the coupling of the usual and chiral charge  currents and their densities by the CMW in Eqs.~\eq{eq:CMW:jVz} and \eq{eq:CMW:jAz}, and by the CVW in Eq.~\eq{eq:CVE:djVA}.

Equations~\eq{eq:mixing2:jAE:3} indicate that, as the wave moves, 
\begin{itemize}

\item[(i)] a local perturbation in the thermal energy $\delta \epsilon$ is converted into a perturbation in the chiral current $\delta {\bs j}_A$;

\item[(ii)] the perturbation in the chiral current $\delta {\bs j}_A$ leads to an excess in the chiral charge density $\delta \rho_A$ the next point along the vector ${\bs \Omega}$; 

\item[(iii)] the perturbation in the chiral charge density $\delta \rho_A$ in turn, induces a perturbation in the energy current $\delta {\bs j}_E$;

\item[(iv)] the energy current $\delta {\bs j}_E$ generates immediately a fluctuation in the thermal energy $\delta \epsilon$ next point etc.

\end{itemize}
The process is very similar to the mutual cyclic conversion of the charge density and chiral charge density waves in the CMW, apart from the fact that in our case the role of the charge density wave is played by the thermal energy (or, heat) wave. Thus, we call this new gapless excitation the Chiral Heat Wave (CHW). 

The fluctuations of energy and chiral currents~\eq{eq:mixing2:jAE:3} propagate along the vorticity vector ${\bs \Omega}$. Therefore we set for convenience ${\bs \Omega} = \Omega {\bs e}_z$, take $\Omega > 0$ for the sake of definiteness, and consider the currents directed along the $z$ axis only. The conservation of the axial charge~\eq{eq:conservation:jVjA:z} and energy~\eq{eq:energy:conservation} give us the relations
\beqn
\begin{split}
\partial_t \delta \rho_A + \partial_z \delta j^z_A = 0\,, \\
\partial_t \delta \epsilon + \partial_z \delta j^z_E = 0\,,
\end{split}
\eeqn
which can now be combined with Eq.~\eq{eq:mixing2:jAE:3}
\beqn
\begin{split}
\partial_t \delta \rho_A +  \frac{{ T} \Omega}{3 c_V} \, \partial_z \delta \epsilon = 0\,,\\
\partial_t \delta \epsilon + \frac{{ T}^2 \Omega}{3 \chi} \partial_z \delta \rho_A = 0 \,.
\end{split}
\label{eq:mixing:EA:2}
\eeqn
The system of equations~\eq{eq:mixing:EA:2} describes a gapless propagation of the coupled energy and chiral density perturbations along the axis of rotation. By combining these first-order equations one gets that both the chiral charge density $\rho_A$ and the energy density $\rho_E \equiv \epsilon$ obey the second-order equations:
\beqn
\begin{split}
\left(\partial_t^2 - v_{\CHW}^2 \partial_z^2\right)  \delta \rho_A(t,z) & = 0\,,\\ 
\left(\partial_t^2 - v_{\CHW}^2 \partial_z^2\right)  \delta \rho_E(t,z) & = 0\,,
\end{split}
\eeqn
where
\beqn
v_{\CHW} = \sqrt{\frac{{T}^3}{c_V \chi}} \frac{\Omega}{3}\,,
\label{eq:v:CHW}
\eeqn
is the velocity of the Chiral Heat Wave. Notice that in our derivation we have implicitly assumed that the usual thermal diffusion and the axial charge relaxation are so slow so that the wave propagates adiabatically. 

Let us now discuss the structure of the CHW in terms of energy and chiral charge densities. According to Eq.~\eq{eq:mixing1:jV:2} the usual charge density, $\rho_V \equiv \rho_R + \rho_L$, does not propagate in this wave so that in the CHW the densities of the right-handed and left-handed fermions are always opposite to each other: $\rho_R = - \rho_L$.  A diagonalization of Eqs.~\eq{eq:mixing:EA:2} indicates that the linear combinations of the energy density and the chiral density,
\beqn
\delta {\cal E}_{\pm}(t,z) = \delta \epsilon(t,z) \pm \sqrt{\frac{c_V T}{\chi}} \delta \rho_A(t,z)\,,
\label{eq:delta:E}
\eeqn
obey, respectively, the following linear equations:
\beqn
\left(\partial_t \pm v_{\CHW} \partial_z\right) \delta {\cal E}_{\pm}(t,z) = 0\,.
\eeqn

\begin{figure}[!thb]
\begin{center}
\includegraphics[scale=0.55,clip=true]{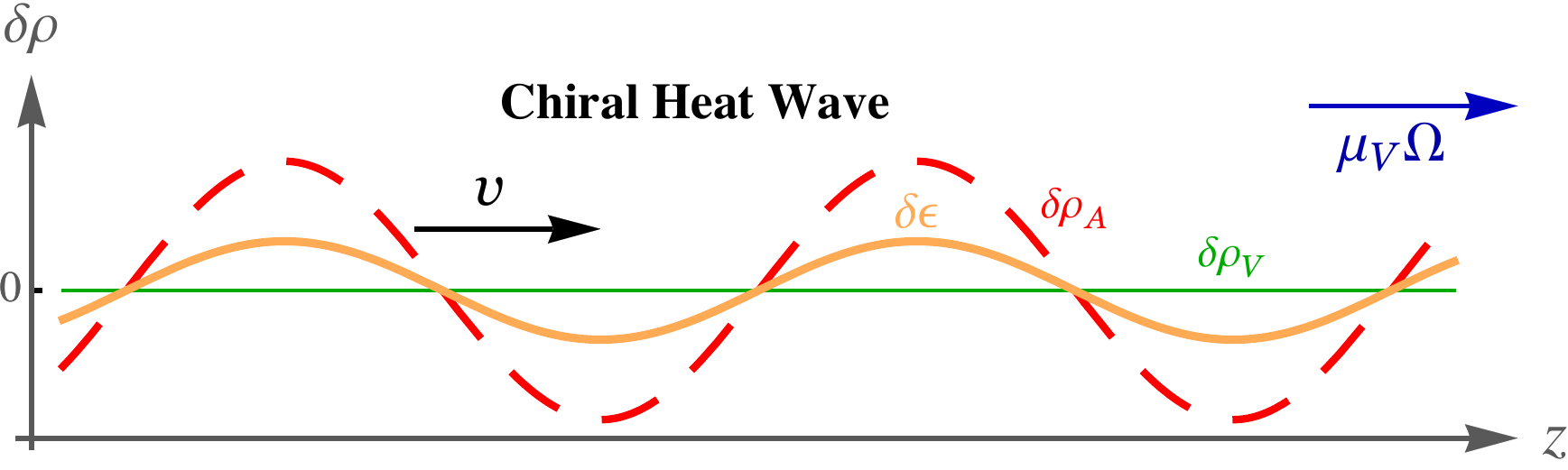}\\[5mm]
\includegraphics[scale=0.55,clip=true]{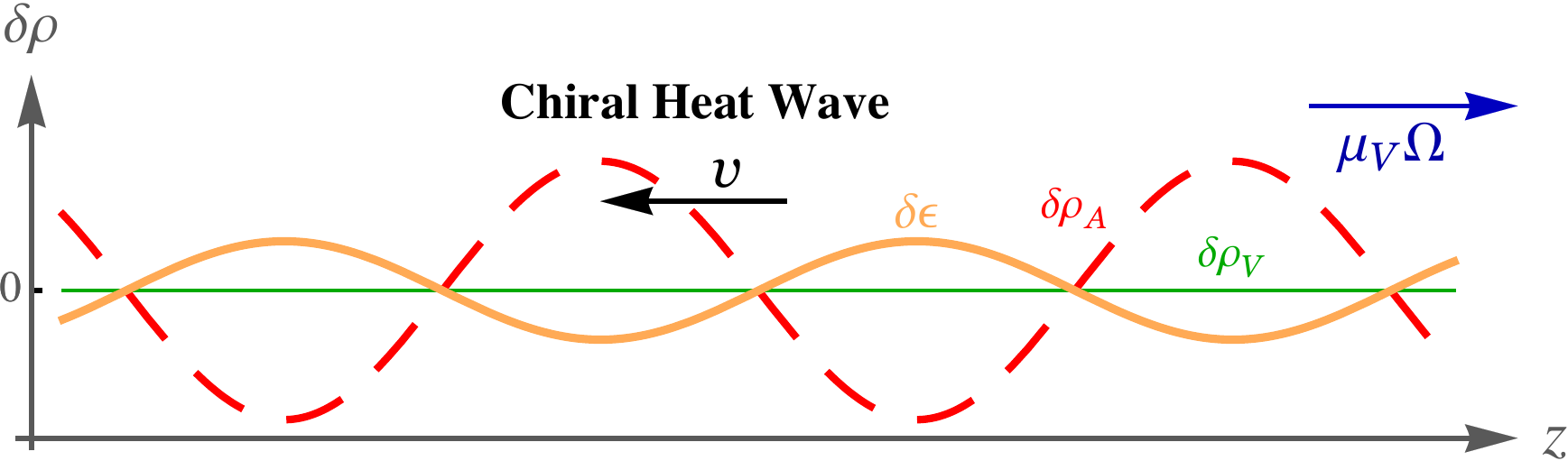}
\end{center}
\caption{Structure of the Chiral Heat Wave in a zero-density ($\mu_V = 0$) hot ($T > 0$) rotating (${\bs \Omega} \neq 0$) chiral medium at vanishing magnetic field (${\bs B} = 0$). The notations follow Fig.~\ref{fig:CMW}.}
\label{fig:CHW}
\end{figure}

The combinations ${\cal E}_+$  and ${\cal E}_-$ represent the pure Chiral Heat Waves, in which the energy and chiral charge densities have, respectively, the same  and opposite mutual signs:
\beqn
\delta \epsilon(t,z) = \pm \sqrt{\frac{c_V { T}}{\chi}} \delta \rho_A(t,z)\,,
\label{eq:modes}
\eeqn
and which propagate, correspondingly, along and against the direction of the vorticity vector. As we have mentioned, the vector charge density component in the pure heat wave is always zero. The structure of the Chiral Heat Wave is illustrated in Fig.~\ref{fig:CHW}.

According to Eq.~\eq{eq:modes}, as temperature ${ T}$ becomes higher the fraction of the thermal energy component increases compared to the chiral component. Any combination of $\delta \epsilon$ and $\delta \rho_A$ perturbations can be expanded into the individual Chiral Heat Waves $\delta {\cal E}_{\pm}$ which are propagating in opposite directions.

The dispersion relations for the  CHWs in the long-wavelength limit reads as follows:
\beqn
\omega^\pm_{\CHW} (k_z) = \pm v_{\CHW} k_z - i D k_z^2 + \dots\,,
\label{eq:omega:CHW}
\eeqn
where upper and lower signs correspond to ${\cal E}_\pm$ modes~\eq{eq:modes}. Here we have also included the dissipative term which was not captured by our linearized analysis (higher-order terms are denoted by the ellipsis). 

In conclusion of this Section we notice that the chiral heat wave bear (due to its thermal nature) a very distant similarity to the ``temperature wave'' (the ``second sound'') of the superfluid Helium. We would like also to notice that all three discussed gapless modes propagate, in general, with different velocities given by Eqs.~\eq{eq:v:CMW}, \eq{eq:v:CVW} and \eq{eq:v:CHW} for magnetic, vortical and heat waves, respectively.

\section{Mixed waves}
\label{sec:mix}

\subsection{Chiral Magnetic-Vortical wave} 
\label{sec:MV}

In this section we point out that the CMW and the CVW should naturally mix with each other and constitute a common wave provided the conditions necessary for simultaneous existence of both these waves (${\bs B} \neq 0$, ${\bs \Omega} \neq 0$ and ${\mu}_V \neq 0$), are satisfied. In this Section we consider the cold medium ($T=0$) in order to exclude the presence of the Chiral Heat Wave.

In the magnetic field background ${\bs B}$ the vector and axial currents of the rotating fluid are given by the linear combination of the CME~\eq{eq:CME:jV}, the CSE~\eq{eq:CSE:jA} and the CVE~\eq{eq:CVE:jVA}:
\beqn
{\bs j}_V & = & \sigma_{V}^{{\cal B}} e {\bs B} + \sigma_{V}^{{\cal V}} {\bs \Omega}\,, 
\label{eq:mixing1:jV}\\
{\bs j}_A & = & \sigma_{A}^{{\cal B}} e {\bs B} + \sigma_{A}^{{\cal V}} {\bs \Omega}\,.
\label{eq:mixing2:jA}
\eeqn
Since the corresponding relations between the fluctuations of the charge densities and currents are linear, we combine Eqs.~\eq{eq:CMW:jVz} and \eq{eq:CMW:jAz} with \eq{eq:CVE:djVA} and get
\beqn
\begin{split}
\delta {\bs j}_V & = \frac{1}{2 \pi^2 \chi} \left( e{\bs B} + 2 { \mu}_V {\bs \Omega} \right) \delta \rho_A \equiv \frac{e {\bs B}^\eff}{2 \pi^2 \chi} \delta \rho_A \,, \\[1mm]
\delta {\bs j}_A & = \frac{1}{2 \pi^2 \chi} \left( e{\bs B} + 2 {\mu}_V {\bs \Omega} \right) \delta \rho_V \equiv \frac{e {\bs B}^\eff}{2 \pi^2 \chi}\delta \rho_V  \,,
\end{split}
\label{eq:mixing1:djVA}
\eeqn
which can again be reduced to Eqs.~\eq{eq:CMW:jVz} and \eq{eq:CMW:jAz} with the substitution ${\bs B} \to {\bs B}^\eff$ where the effective magnetic field~is:
\beqn
e {\bs B}^\eff = e {\bs B} + 2 {\mu}_V {\bs \Omega} \,.
\label{eq:B:eff}
\eeqn

From Eq.~\eq{eq:mixing1:djVA} we see that the CMW and CVW couple to each other and constitute one common wave which propagates with the velocity
\beqn
v_{\MV} & = & \frac{e B^\eff}{2 \pi^2 \chi} = \frac{|e {\bs B} + 2 {\mu}_V {\bs \Omega}|}{2 \pi^2 \chi} 
\equiv \sqrt{v_\CMW^2 + v_\CVW^2 + 2 v_\CMW v_\CVW \cos \varphi({{\bs B},{\bs \Omega}})}\,, \qquad
\label{eq:v:MV}
\eeqn
along the axis given by the direction of the vector ${\bs B}^\eff$. The velocity of the common wave depends on the angle $\varphi({{\bs B},{\bs \Omega}})$ between the magnetic field and the axis of rotation. The subscript ``MV'' in Eq.~\eq{eq:v:CVW} refers to the mixing of the Chiral {\underline M}agnetic and {\underline V}ortical Waves.

Thus, the mix of the magnetic and vortical waves leads to the change of the propagation direction of the common mixed wave as compared to the directions of the individual waves. Indeed, the individual CMW wave in the absence of the CVW would propagate along the magnetic field ${\bs B}$ while the individual CVW in the absence of the CMW would propagate along the rotation velocity ${\bs \Omega}$. In the a rotating dense system subjected to magnetic field the CMW and CVW always mix and form a common wave which propagates only along the common vector~\eq{eq:B:eff}. 

The velocity of the common wave is also changed~\eq{eq:v:CVW} compared to the velocities of the pure CMW~\eq{eq:v:CMW} and CVW~\eq{eq:v:CVW}. The charge-density structure of the waves remain, however, the same: the right-handed modes (with the same vector and axial densities) propagate along the effective magnetic field~\eq{eq:B:eff} while the left-handed modes (with opposite vector and axial densities) propagate opposite to this field.

The energy density component of the mixed magnetic-vortical wave can be easily found from Eqs.~\eq{eq:jE}, \eq{eq:sigma:EB} and \eq{eq:sigma:EV}:
\beqn
\delta {\bs j}_E = \mu_V \frac{e {\bs B}^\eff}{2 \pi^2 \chi} \delta \rho_A \equiv \mu_V \delta {\bs j}_V \,,
\label{eq:energy:slave:wave}
\eeqn
where we have also used the first relation in Eq.~\eq{eq:mixing1:djVA}. Equation~\eq{eq:energy:slave:wave} demonstrates that the energy component of the common magnetic-vortical wave is, basically, the a mass ``slave'' wave which is tightly bound to the vector charge density wave. In particular, $\delta \epsilon (t,{\bs x}) = \mu_V \delta \rho_V  (t,{\bs x})$. Thus, the qualitative structure of the mixed Chiral Magnetic/Vortical Wave is the same as the structure of the Chiral Vortical Wave (shown in Fig.~\ref{fig:CVW}) with the reservation that the direction of the mixed wave propagation is collinear to the axis of the effective field~\eq{eq:B:eff}.

One of the possible environments where both CMW and CVW may realize is a noncentral heavy-ion collision. The created fireball of quark-gluon plasma is, basically, a rotating fluid of light fermions in a background of strong magnetic field (the latter is created by the ion constituents and by the products of their collision). In this case the vectors ${\bs \Omega}$ and ${\bs B}$ are co-aligned with each other and the velocity of the common Magnetic-Vortical wave is a sum of the velocities of the individual CMW and CVW~\eq{eq:v:MV}: 
\beqn
v_{\MV} = v_\CMW \pm v_\CVW \qquad \mbox{for} \quad {\bs B} \| {\bs \Omega}\,,
\label{eq:v:MV:parallel}
\eeqn
where the upper (lower) sign corresponds to parallel (antiparallel) orientation of $e {\bs B}$ and $\mu_V {\bs \Omega}$.\footnote{Here we have explicitly included the prefactors $e$ and $\mu_V$ since they may take negative values.}

Notice that for the specific relation between the magnetic field and the angular velocity, 
\beqn
{\bs B} = - \frac{2 { \mu}_V}{e} {\bs \Omega} \,,
\label{eq:B:special}
\eeqn
the chiral waves do not propagate in the system at all. Indeed, in the considered environment (${\bs B} \neq 0$, ${\bs \Omega} \neq 0$ and ${\mu}_V \neq 0$) the pure CMW and the pure CVW do not exist alone so that the vector and axial density waves propagate only in the form of the mixed Chiral Magnetic/Vortical Wave. For the strength of magnetic field~\eq{eq:B:special} the velocity of the mixed wave is zero, so that no wave propagation occurs.

\subsection{Chiral Vortical-Heat Wave}
\label{sec:VH}

\subsubsection{Three solutions for dispersions}

In the previous section we have shown that the Chiral Heat Wave emerges in a rotating finite-temperature fluid at zero chemical potential in the absence of magnetic field. Let us lift a bit our restrictions and consider the same fluid but with a small
nonzero chemical potential, $\mu_V \neq 0$.\footnote{In order to keep our analysis simple, we mostly consider slowly rotating systems at small chemical potential ($\Omega \ll T$, $\mu_V \ll T$), so that $O(\Omega^2)$ and $O(\mu^2_V)$ terms in energy density~\eq{eq:epsilon} can be neglected.} 
We already know that the Chiral Vortical Wave should appear in this environment as a gapless excitation. These two waves should mix with each other because the CHW propagates as the coupled energy and axial (chiral) charge density wave, while the CVW couples the vector and axial charge densities. Since the CHW and CVW have one common axial channel they should inevitably mix at a finite density. 

Following our previous tactics we use the full system of equations \eq{eq:mixing1:jV}, \eq{eq:mixing2:jA} and \eq{eq:jE} along with Eqs.~\eq{eq:delta:q:a} and \eq{eq:delta:epsilon} to derive relations between the fluctuations of currents~${\bs J}$ and densities~$J^0 \equiv Q$:
\beqn
\delta {\vec {\bs J}} = {\bs \Omega}  {\hat M} \delta {\vec J}^{\,0}\,,
\label{eq:J:hatM}
\eeqn
where we used the vector notations in the charge, axial charge and energy space:
\beqn
{\vec J}^\mu \equiv ({\vec Q}, {\vec {\bs J}}) = 
\left(
\begin{array}{c}
j^\mu_V \\[1mm]
j^\mu_A \\[1mm]
j^\mu_E 
\end{array}
\right)\,,
\label{eq:vec:QJ}
\eeqn
with the matrix
\beqn
{\hat M} =  
\left(
\begin{array}{ccc}
    0   &    \frac{{\mu}_V}{\pi^2 \chi}   &    0    \\
     \frac{{\mu}_V}{\pi^2 \chi}   &   0    &   \frac{{ T}}{3 c_V}    \\
    0   &    \frac{{ T}^2}{3 \chi} + \frac{\mu_V^2}{\pi^2 \chi}   &   0     
\end{array}
\right) \Omega.
\label{eq:hat:M}
\eeqn

Using the vector form for the conservation laws of vector and axial charges~\eq{eq:conservation:jVjA:z} and that for the energy~\eq{eq:energy:conservation},
\beqn
\partial_\mu \delta J^\mu = 0\,,
\eeqn
one gets from Eq.~\eq{eq:J:hatM}:
\beqn
\left( \unity \, \partial_t +  {\hat M} \partial_z\right) \delta {\vec Q}(t,z) = 0\,.
\label{eq:diff:1}
\eeqn

It is convenient to use the plane-wave ansatz in Eq.~\eq{eq:diff:1},
\beqn
{\vec Q}(t,z) = {\vec C}_0 e^{- i \omega t + i k_x z}\,.
\label{eq:vecQ}
\eeqn
It appears that there are three solutions for the dispersion relation:
\beqn
\omega_\VH^\pm({\bs k}) = \pm v_{\VH} k_z\,,
\label{eq:omega:kz}
\eeqn
and
\beqn
\omega({\bs k}) = 0\,.
\label{eq:omega:0}
\eeqn
In Eq.~\eq{eq:omega:kz} the subscript ``VH'' stands for the mixed Chiral {\underline V}ortical--{\underline H}eat Wave.

Below we discuss the linear dispersion laws~\eq{eq:omega:kz} corresponding to the mixing between heat and vortical waves. The special (non propagating) zero-frequency solution~\eq{eq:omega:0} is discussed in Section~\ref{sec:DHS:1}.

\subsubsection{Propagating mode: vector-axial-energy wave at $T \neq 0$}

The dispersion~\eq{eq:omega:kz} corresponds to the velocity of the common, Chiral Vortical and Chiral Heat, gapless mode:
\beqn
v_{\VH} & = & \Omega \sqrt{\frac{{\mu}_V^2}{\pi^4 \chi^2} + \frac{{ T}}{3 c_V \chi} \left(\frac{{ T}^2}{3} + \frac{{ \mu}_V^2}{\pi^2} \right)} 
\equiv \sqrt{v_\CVW^2 + \left[ 1 + 3 {\left(\frac{\mu_V}{\pi T} \right)}^2 \right]v_\CHW^2}\,.
\label{eq:v:HV}
\label{eq:v:VH}
\eeqn
Thus, in the rotating dense hot fluid the CVW and the CHW form one vortical-heat wave propagating with velocity~\eq{eq:v:HV} along the vorticity vector ${\bs \Omega}$. Notice that the velocity of the mixed CVW/CHW excitation~\eq{eq:v:HV} is higher compared to the individual velocities of the pure CVW~\eq{eq:v:CVW} and CHW~\eq{eq:v:CHW}.

\begin{figure}[!thb]
\begin{center}
\includegraphics[scale=0.6,clip=true]{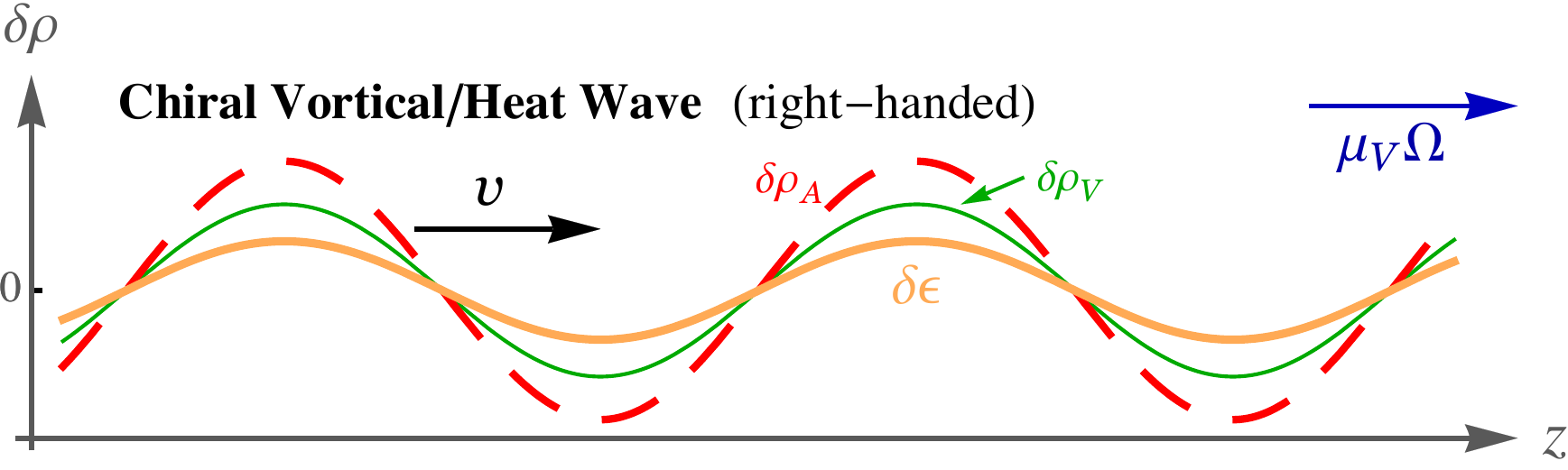}\\[5mm]
\includegraphics[scale=0.6,clip=true]{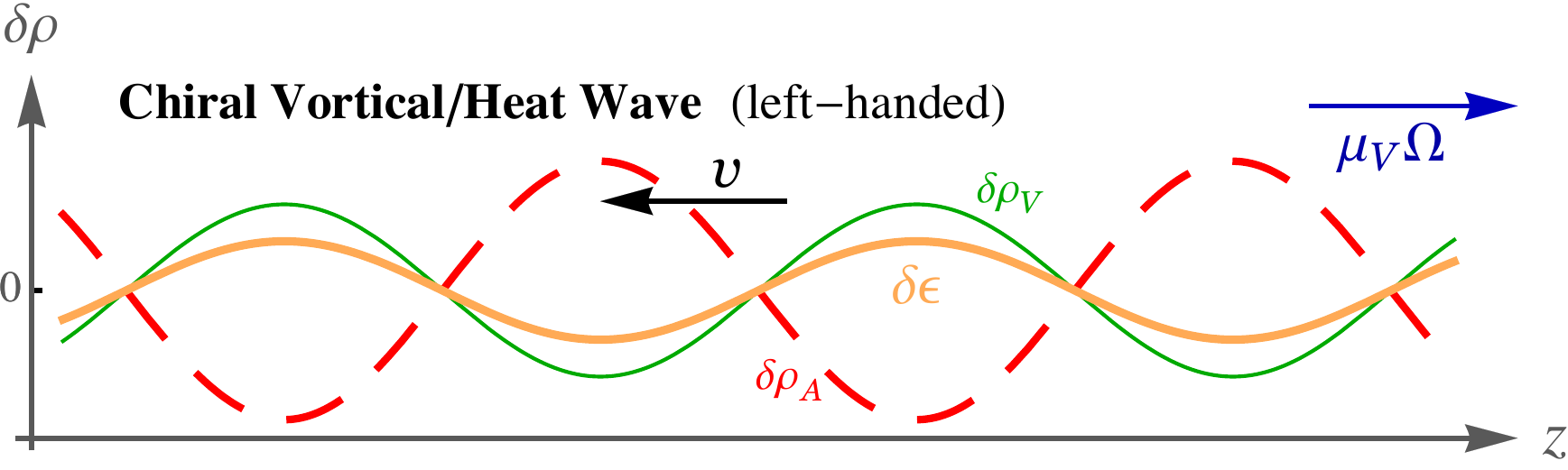}
\end{center}
\caption{Structure of the mixed Chiral Vortical/Heat Wave in a finite-density ($\mu_V > 0$) hot ($T > 0$) rotating (${\bs \Omega} \neq 0$) chiral medium at vanishing magnetic field (${\bs B} = 0$). The notations and the $\mu_V < 0$ case are described in the caption of Fig.~\ref{fig:CVW}.}
\label{fig:VH}
\end{figure}

The vector, axial and energy densities inside the mixed CVW/CHW are interrelated:
\beqn
\delta \rho_A(t,z) & = & \pm \frac{v_\VH}{v_\CVW} \delta \rho_V(t,z)\,, 
\label{eq:qA:qV:HV}\\
\delta \epsilon(t,z) & = & \left({ \mu}_V + \frac{\pi^2 { T}^2}{3 { \mu}_V}\right) \delta \rho_V(t,z)\,, 
\label{eq:epsilon:qV:HV}
\eeqn
where the upper and lower signs correspond to those of Eq.~\eq{eq:omega:kz}, the common vortical-heat wave velocity $v_\VH$ is given in Eq.~\eq{eq:v:HV} and the pure CVW velocity $v_\CVW$ velocity can be found in Eq.~\eq{eq:v:CVW}. Equations~\eq{eq:qA:qV:HV} and \eq{eq:epsilon:qV:HV} correspond to the eigenvectors~\eq{eq:vecQ} of Eq.~\eq{eq:diff:1} with the dispersion relations~\eq{eq:omega:kz}. As expected, the mixed heat/vortical wave is composed of all three vector, axial and energy waves (shown in Fig.~\ref{fig:VH}).

The relation between the vector and axial charges in the mixed wave~\eq{eq:qA:qV:HV} can also be rewritten as follows:
\beqn
\delta \rho_A(t,z) & = & \pm \sqrt{1 + \frac{\pi^2 \chi { T}}{3 c_V} \left(1 + \frac{\pi^2 { T}^2}{3 { \mu}_V^2}\right)} \, \delta \rho_V(t,z)\,. \qquad
\label{eq:qA:qV:HV:simple}
\eeqn
Thus, in the mixed wave the axial charge density is always larger than the vector (usual) charge density. 

The $\omega^+$ modes -- which are primarily composed of the right-handed density wave with admixture of the left-handed wave -- are propagating along with the direction of the rotational velocity (vorticity) $\bs \Omega$, while the $\omega^-$ modes are predominantly left-handed densities which propagate opposite to $\bs \Omega$. Both these modes are coherent with energy density waves which have the same sign as the charge density~\eq{eq:epsilon:qV:HV}  as it is shown in Fig.~\ref{fig:VH}.

\subsubsection{Zero temperature: chiral vortical and mass waves}

In the low-temperature limit ${ T} \to 0$ the pure CHW does not propagate~\eq{eq:v:CHW} as its velocity vanishes, $v_\CHW(T=0) = 0$. In this limit the $\omega^+$ and $\omega^-$ mixed modes~\eq{eq:omega:kz} become, respectively, the pure right-handed and pure left-handed modes carried by the CVW. Notice that even in this limit -- when the Chiral Heat Wave is absent -- the energy density in the CVW is nonzero due to the presence of the chemical potential ${ \mu}_V \neq 0$. Basically, the Chiral Vortical Wave induces a mass wave in the cold matter~\eq{eq:epsilon:qV:HV}. The mass wave is propagating without mass transfer, at least in the linear approximation. Moreover, according to Eq.~\eq{eq:v:HV}, the emerging mass waves are not affecting the velocity of propagation of the CVW provided the temperature is sufficiently low, so that the following conditions are both satisfied:
\beqn
T \ll {\left( \frac{c_V \mu_V^2}{\pi^2 \chi} \right)}^{\frac{1}{3}}\,, 
\qquad
T \ll \frac{c_V}{\pi \chi}\,.
\label{eq:CVW:T:conditions}
\eeqn

The charge and energy density content inside the CMW can be easily read from Eqs.~\eq{eq:qA:qV:HV} and \eq{eq:epsilon:qV:HV}:
\beqn
\delta \rho_A
& = & \pm \delta \rho_V
\,, 
\qquad
\delta \epsilon
= \mu_V \delta \rho_V
\,.
\label{eq:epsilon:qV:mass:V}
\eeqn
We recover the result of Ref.~\cite{ref:CVW} which shows the existence of the pure right-handed and left-handed waves which are propagating in opposite directions. In addition, we have found that each of these density waves induces the mass wave which is proportional to the chemical potential $\mu_V$:
\beqn
L: \ \left\{
\begin{array}{l}
v_L = - v_{\CVW}\,,\\
\delta \rho_L \neq 0\,,\\ 
\delta \rho_R = 0\,, \\
\delta \epsilon = \mu_V \delta \rho_L\,,
\end{array}
\right.
\qquad\qquad
R: \ \left\{
\begin{array}{l}
v_R = + v_{\CVW}\,,\\
\delta \rho_L = 0\,,\\ 
\delta \rho_R \neq 0\,, \\
\delta \epsilon = \mu_V \delta \rho_R\,.
\end{array}
\right.
\qquad
\eeqn

Summarizing, at sufficiently low temperatures~\eq{eq:CVW:T:conditions}, neither the chirality nor the velocity of the CVWs is affected by the presence of the energy wave induced by the mixed gauge-gravitational anomaly. However, at higher temperatures the propagating CVW is always accompanied by the thermal (``heat'') energy density wave~\eq{eq:epsilon:qV:HV}. The velocity of the mixed vortical-heat wave is higher compared to the velocity of the original vortical wave. Moreover, the mixed chiral vortical-heat wave is neither right-handed or left-handed~\eq{eq:qA:qV:HV} contrary to the pure CVW.

\subsection{Chiral Magnetic-Heat Wave}
\label{sec:MH}

Now let us consider the case of rotating (${\bs \Omega} \neq 0$) finite-temperature ($T\neq 0$) neutral (${ \mu}_V = 0$) fluid in external magnetic field (${\bs B} \neq 0$). Since the chemical potential is absent then the chiral vortical wave does not exist. The propagation of the chiral magnetic and heat waves is described by the following relation between the charge and density fluctuations:
\beqn
\delta {\bs j}_V & = & \frac{ e {\bs B}}{2 \pi^2 \chi} \delta \rho_A\,, 
\label{eq:mixing1:jV:22}\\
\delta {\bs j}_A & = & \frac{ e {\bs B}}{2 \pi^2 \chi} \delta \rho_V + \frac{T {\bs \Omega}}{3 c_V} \delta \epsilon\,, 
\label{eq:mixing2:jA:22}\\
\delta {\bs j}_E & = & \frac{T^2 {\bs \Omega}}{3 \chi} \delta \rho_A \,.
\label{eq:jE:22}
\eeqn
With the help of the conservation laws~\eq{eq:cons:jV}, \eq{eq:cons:jA} and \eq{eq:energy:conservation}, these equations may be cast into the system of linear equations:
\beqn
\left( \unity \, \partial_t + ({\hat {\bs M}} \cdot {\bs \partial})\right) \delta {\vec Q}(t,z) = 0\,,
\label{eq:diff:2}
\eeqn
where the charge vector ${\vec Q}$ is given in Eq.~\eq{eq:vec:QJ} and
\beqn
{\hat {\bs M}} =  
\left(
\begin{array}{ccc}
    0   &    \frac{e {\bs B}}{2 \pi^2 \chi}   &    0    \\
     \frac{e {\bs B}}{2 \pi^2 \chi}   &   0    &   \frac{{ T} {\bs \Omega}}{3 c_V}    \\
    0   &    \frac{{ T}^2 {\bs \Omega}}{3 \chi}  &   0     
\end{array}
\right).
\label{eq:hat:M:2}
\eeqn
The dispersion relation for the density waves in the system reads as follows:
\beqn
\left(V_{ij} k^i k^j - \omega^2 \right) \omega = 0\,,
\label{eq:Vij:omega}
\eeqn
where the matrix in the momentum space
\beqn
V_{ij}({\bs B},{\bs \Omega}) = v_{\CMW}^2 (B) \, e_i^{B} e_j^{B} + v_{\CHW}^2 (\Omega) \, e^{\Omega}_i e^{\Omega}_j, \qquad
\label{eq:Vij:def}
\eeqn
depends on two unit vectors directed along the magnetic field and the angular velocity, respectively:
\beqn
{\bs e}^B = \frac{{\bs B}}{B}\,, 
\qquad
{\bs e}^\Omega = \frac{{\bs \Omega}}{\Omega}\,.
\label{eq:B:Omega:e}
\eeqn
In Eq.~\eq{eq:Vij:def} the velocities of the pure magnetic and pure heat waves are given in Eqs.~\eq{eq:v:CMW} and \eq{eq:v:CHW}, respectively.

There are three branches of solutions of Eq.~\eq{eq:Vij:omega} for the dispersion relations:
\beqn
\omega = \pm \sqrt{V_{ij} k^i k^j}\,,
\label{eq:omega:Vij:MH}
\eeqn
and
\beqn
\omega = 0\,.
\label{eq:omega:0:MH}
\eeqn
The trivial solution~\eq{eq:omega:0:MH} is considered in Sect.~\ref{sec:DHS:2} below. 

In order to deal with the nontrivial dispersion relation~\eq{eq:omega:Vij:MH} we simplify (without loss of generality) our calculations by setting the rotation vector $\bs \Omega$ along the $z$-axis and turning the vector of magnetic field ${\bs B}$ in the $xz$-plane:
\beqn
{\bs B} = (B \sin \varphi, 0, B \cos \varphi)\,, \qquad 
{\bs \Omega} = (0,0,\Omega)\,, \qquad
\label{eq:varphi:B}
\eeqn
where $\varphi \equiv \varphi ({\bs B}, {\bs \Omega})$ is the angle between the magnetic field ${\bs B}$ and the rotation axis ${\bs \Omega}$.

The eigensystem of the matrix~\eq{eq:Vij:def} is was follows:
\beqn
{\widehat V}  {\bs e}_a = \lambda_a {\bs e}_a \,, \qquad a = \pm, y\,,
\eeqn
where the trivial eigenvalue $\lambda_y = 0$ corresponds to the (unit-length) eigenvector ${\bs e}_y$ along the $y$ axis, which is orthogonal to both ${\bs \Omega}$ and ${\bs B}$. The nonzero eigenvalues, $\lambda_\pm  \equiv v_{\MH,\pm}^2$, can be expressed via the following quantities:
\beqn
v_{\MH,\pm}^2 & = & \frac{1}{2} \biggl( v_{\CMW}^2 + v_{\CHW}^2  \pm \sqrt{\left( v_{\CMW}^2 + v_{\CHW}^2\right)^2 - 4 v_{\CMW}^2 v_{\CHW}^2 \sin^2\varphi}\, \biggr)\,.
\label{eq:v:pm} 
\eeqn
As we will see below $v_{\MH,\pm}$ are two principal velocities of the coupled Chiral {\underline M}agnetic and Chiral {\underline H}eat Waves (denoted by the subscript ``MH'').

The eigenvectors corresponding to the eigenvalues~\eq{eq:v:pm} are located in the $xz$ plane,
\beqn
\begin{split}
{\bs e}_+ & = (\sin\theta,0, \cos\theta) \,, \\
{\bs e}_- & = (-\cos\theta,0, \sin\theta)\,,
\end{split}
\label{eq:n:plus:minus}
\eeqn
where the angle $\theta$ depends on the strength and mutual orientation of the magnetic field and the rotation velocity:
\beqn
\theta({\bs B},{\bs \Omega}) = \arctan \left[\left(1 - \frac{v^2_{\CHW}({\Omega})}{v_{\MH,+}^2({\bs B},{\bs \Omega})} \right) \tan \varphi({\bs B}, {\bs \Omega}) \right]\!. \qquad
\label{eq:theta:MH}
\eeqn
The eigenvectors ${\bs e}_y$ and ${\bs e}_\pm$ form the orthonormal system, ${\bs e}_a {\bs e}_b = \delta_{ab}$ with $a,b=\pm,y$.

It is convenient to re-express the momentum vector ${\bs k}$ in the orthogonal basis $({\bs e}_+,{\bs e}_-,{\bs e}_y)$, 
\beqn
{\bs k} = \sum_{a=\pm,z} k_a {\bs e}_a\,,
\eeqn
where 
\beqn
\begin{split}
k_+ & = k_x \sin \theta + k_z \cos \theta\,, \\
k_- & = -k_x \cos \theta + k_z \sin \theta\,,
\end{split}
\eeqn
and the momentum component $k_y$ remains unrotated. Then one gets the following dispersion relation from Eq.~\eq{eq:Vij:omega}
\beqn
\omega({\bs k}) = \pm \sqrt{v_{\MH,+}^2 k_+^2 + v_{\MH,-}^2 k_-^2}\,,
\eeqn
where the velocities $v_{\MH,\pm}$ are defined in Eq.~\eq{eq:v:pm}.

The velocity of the wave propagation is
\beqn
{\bs v}({\bs k}) \equiv \frac{\partial \omega({\bs k})}{\partial {\bs k}} = \frac{1}{\omega} {\widehat V} {\bs k}\,,
\eeqn
where the matrix ${\hat V}$, which operates in the coordinate space, is defined in Eq.~\eq{eq:Vij:def}. One gets explicitly:
\beqn
{\bs v} = \frac{v_{\MH,+}^2 k_+ {\bs e}_+ + v_{\MH,-}^2 k_- {\bs e}_-}{\sqrt{v_{\MH,+}^2 k_+^2 + v_{\MH,-}^2 k_-^2}}\,.
\label{eq:vec:v:MH}
\eeqn

First, we conclude from Eq.~\eq{eq:vec:v:MH} that for non-collinear ${\bs \Omega}$ and ${\bs B}$, the propagation of the mixed magnetic/heat wave is not constrained to a singe vector contrary to the cases of the pure magnetic wave or pure heat wave. The mixed wave may propagate in the whole $xz$ plane spanned on the vectors ${\bs \Omega}$ and ${\bs B}$. Second, Eq.~\eq{eq:vec:v:MH} indicates that the wave vector ${\bs k}$ and the velocity vector ${\bs v}$, if even they are constrained to belong to the common $xz$ plane, are not parallel unless
\begin{itemize}

\item[(i)] The wave vector ${\bs k}$ is directed along one of the eigenvectors ${\bs e}_\pm$ (so that either $k_+$ or $k_-$ is zero). Then the wave propagates along the vectors ${\bs e}_+$ and ${\bs e}_-$ with the velocities $v_{\MH,+}$ and $v_{\MH,-}$ [given in Eq.~\eq{eq:v:pm}], respectively;

\item[(ii)] One of the $v_{\MH,\pm}$ velocities vanishes (so that $\sin \varphi = 0$, implying that the rotation axis and the magnetic field are collinear to each other, ${\bs \Omega} \| {\bs B}$); 

\item[(iii]) If $v_{\MH,+} = v_{\MH,-}$. This is possible if two conditions are satisfied~\eq{eq:v:pm}: the rotation axis and the magnetic field should be perpendicular to each other ${\bs \Omega} \perp {\bs B}$, so that $\varphi = \pm \pi/2$, and the both CMW and CHW velocities should be equal, $v_\CMW = v_\CHW$. 

\end{itemize}

Thus, in general, the direction of the phase velocity, given by the wave vector ${\bs k}$ and the direction of the group velocity, given by the vector ${\bs v}$ do not coincide with each other, which is not unexpected given the anisotropic nature of the medium. In order to illustrate this anisotropy, let us consider the the wave vector ${\bs k}$ directed along the axis of rotation ${\bs \Omega}$ so that ${\bs k} \| {\bs \Omega} \| {\bs e}_z$. Then from Eqs.~\eq{eq:v:pm}. \eq{eq:theta:MH} and \eq{eq:vec:v:MH} we conclude that the mixed wave propagates with the following velocity 
\beqn
{\bs v}_0 = {v}_{0} (\sin \vartheta, 0, \cos \vartheta)\,,
\eeqn
where 
\beqn
{v}_{0}
& =  & \sqrt{\frac{(v_\CHW^2 + v_\CMW^2 \cos^2\varphi)^2 + \frac{1}{4} v_\CMW^4 \sin^2 2 \varphi}{v_\CHW^2 + v_\CMW^2 \cos^2\varphi}}\,, \qquad \\
\vartheta & = & \arctan \frac{v_\CMW^2 \sin \varphi \cos \varphi}{v_\CHW^2 + v_\CMW^2 \cos^2\varphi}\,.
\label{eq:vartheta:B}
\eeqn

\begin{figure}[!thb]
\begin{center}
\includegraphics[scale=0.6,clip=true]{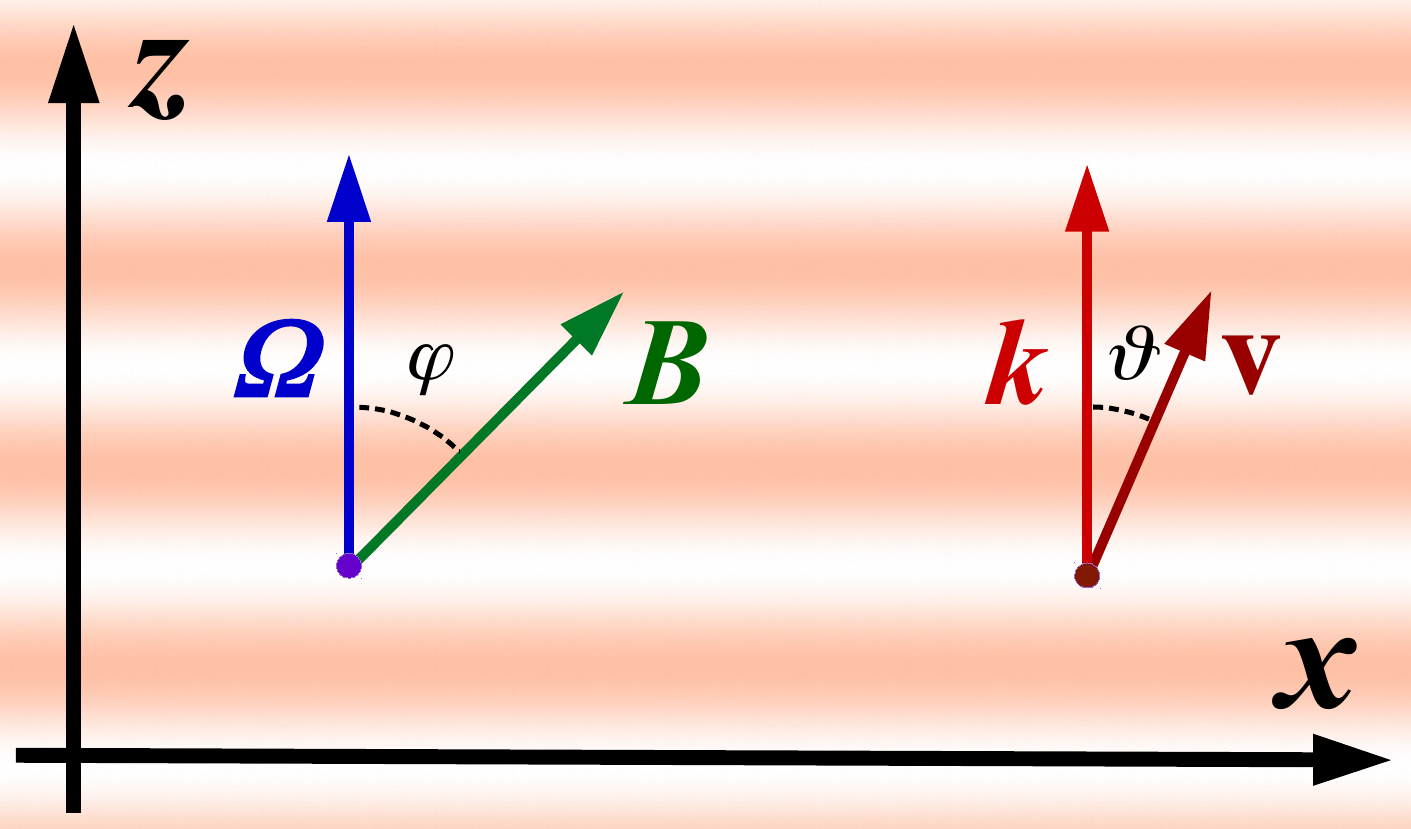}
\end{center}
\caption{Illustration of the anisotropic propagation of the mixed Chiral Magnetic/Heat Wave in the plane spanned on the vectors of the magnetic field ${\bs B}$ and the rotation axis ${\bs \Omega}$. If ${\bs B}$ and ${\bs \Omega}$ are not collinear, then the direction of the wave vector ${\bs k}$ and the direction of the wave velocity ${\bs v}$ may not coincide with each other. The angles $\varphi$ and $\vartheta$ are defined in Eqs.~\eq{eq:varphi:B} and \eq{eq:vartheta:B}, respectively.}
\label{fig:waves}
\end{figure}

For example, if the angle between the magnetic field and the rotation axis is $\varphi = \pi/4 = 45^\circ$, and the parameters of the gas are chosen in such a way that the velocities for the pure chiral magnetic~\eq{eq:v:CMW} and heat~\eq{eq:v:CHW} waves are the same, $v = v_\CMW = v_\CHW$, then the angle between the phase and group velocity is $\vartheta \approx 0.32 \approx 18^\circ$ while the velocity of the mixed wave is greater than the velocities of any of its pure constituents: 
$v_{\MH}^\parallel= \sqrt{5/3} v \approx 1.3 v$. We visualize this anisotropic effect in Fig.~\ref{fig:waves}.

\vskip 3mm

The coupled chiral magnetic-heat wave carries the vector, charge and energy densities similarly to the coupled vortex-heat wave. 

Let us consider the case when the magnetic field and angular velocity are collinear to each other, ${\bs B}\| {\bs \Omega} \| {\bs e}_z$. This case is relevant to noncentral heavy-ion collisions. Setting $\varphi=0$ in Eq.~\eq{eq:v:pm} we obtain the velocity of the coupled CMW/CHW excitation:
\beqn
v^\|_{\MH} = \sqrt{v_{\CMW}^2 + v_{\CHW}^2} \equiv \sqrt{\frac{e^2 B^2}{4 \pi^4 \chi^2} + \frac{{ T}^3 \Omega^2}{9 c_V \chi}}\,,
\label{eq:v:MH}
\eeqn
where we have used Eqs.~\eq{eq:v:CMW} and \eq{eq:v:CHW}. In this simple case the mixed wave propagates along the common axis of ${\bs B}$ and ${\bs \Omega}$, and the directions of the wave and velocity vectors are collinear to each other, ${\bs B} \| {\bs \Omega} \| {\bs k} \| {\bs v}$.

Finally, we mention that in the low-temperature limit the CHW ceases to exist while the CMW wave remains unaffected. The fluctuations in thermal energy decouple from the vector and axial fluctuations according to Eqs.~\eq{eq:mixing1:jV:22}, \eq{eq:mixing2:jA:22} and \eq{eq:jE:22}. Thus, the mixed magnetic-heat wave becomes the pure Chiral Magnetic Wave at $T \to 0$.

\subsection{Chiral Magnetic-Vortical-Heat Wave}
\label{sec:mix:MVH}

Finally, let us assume that all discussed ingredients are present: we consider hot ($T \neq 0$) rotating (${\bs \Omega} \neq 0$) finite-density (${ \mu}_V \neq 0$) fluid subjected to the external magnetic field ${\bs B} \neq 0$. The fluctuations of currents $\delta {\bs j}_a$ and the densities $\delta \rho_a$ (with $a=V,A,E$) are now related as follows:
\beqn
\delta {\bs j}_V & = & \frac{\delta \rho_A}{2 \pi^2 \chi} e {\bs B} + \frac{\mu_V \delta \rho_A}{\pi^2 \chi} {\bs \Omega}\,, 
\label{eq:mixing1:jV:3}\\
\delta {\bs j}_A & = & \frac{\delta \rho_V}{2 \pi^2 \chi} e {\bs B} + \left(\frac{T \delta \epsilon}{3 c_V} + \frac{\mu_V \delta \rho_V}{\pi^2 \chi} \right) {\bs \Omega}\,, 
\label{eq:mixing2:jA:3}\\
\delta {\bs j}_E & = & \frac{\mu_V \delta \rho_A}{2\pi^2 \chi} e {\bs B} + \frac{\delta \rho_A}{\chi} \left(\frac{\mu_V^2}{\pi^2} + \frac{T^2}{3} \right) {\bs \Omega}\,,
\label{eq:jE:3}
\eeqn
where we have used Eqs.~\eq{eq:CME:jV}, \eq{eq:CSE:jA}, \eq{eq:currents}, \eq{eq:CVE:jVA}, \eq{eq:sigma:CVE}, \eq{eq:jE}, \eq{eq:sigma:EB} and \eq{eq:sigma:EV}. Then the conservation laws~\eq{eq:cons:jV}, \eq{eq:cons:jA} and \eq{eq:energy:conservation} give us the familiar relation~\eq{eq:diff:2} with, however, the following new matrix:
\beqn
{\hat {\bs M}} =  
\left(
\begin{array}{ccc}
    0   &    \frac{e {\bs B} + 2 \mu_V {\bs \Omega}}{2 \pi^2 \chi} &    0    \\
     \frac{e {\bs B} + 2 \mu_V {\bs \Omega}}{2 \pi^2 \chi}   &   0    &   \frac{{ T} {\bs \Omega}}{3 c_V}    \\
    0   &  \frac{(e {\bs B} + 2 \mu_V {\bs \Omega})\mu_V }{2 \pi^2 \chi} +  \frac{{ T}^2 {\bs \Omega}}{3 \chi}  &   0     
\end{array}
\right). \quad
\label{eq:hat:M:3}
\eeqn
The magnetic field enters Eq.~\eq{eq:hat:M:3} only via its linear combination ${\bs B}^\eff$ with the angular velocity vector~\eq{eq:B:eff}. As we have seen, this is a feature of the CMW/CVW mixing. Moreover, the structure of the matrix~\eq{eq:hat:M:3} coincides with the matrix~\eq{eq:hat:M:2} which describes the mixing of the chiral magnetic and chiral heat waves. 

Exploring this analogy further, we notice that the excitation spectrum consists of the trivial branch $\omega \equiv 0$ and two gapless waves~\eq{eq:omega:Vij:MH} with
\beqn
V_{ij} & = & v^2_{\MV} e^\eff_i e^\eff_j + v_{\CHW}^2 e_i^{\Omega} e_j^{\Omega} 
+ \gamma v_{\MV} v_{\CHW} \left( e^\eff_i e_j^{\Omega} + e^{\Omega}_i e_j^\eff \right). \qquad
\label{eq:Vij:2}
\eeqn
Here $v_{\MV}$ is the velocity of the mixed Chiral Magnetic/Vortex Wave~\eq{eq:v:MV}, ${\bs e}^\Omega$ is defined in Eq.~\eq{eq:B:Omega:e}, ${\bs e}^\eff$ is the unit vector in the direction of the effective magnetic field~\eq{eq:B:eff}:
\beqn
{\bs e}^\eff = \frac{{\bs B}^{\eff}}{|{\bs B}^{\eff}|} \equiv \frac{e {\bs B} + 2 { \mu}_V {\bs \Omega}}{|e {\bs B} + 2 { \mu}_V {\bs \Omega}|}\,,
\eeqn
and
\beqn
\gamma = \frac{1}{2} \frac{v_\CHW}{v_\CVW}  {\left(\frac{\mu_V}{\pi T}\right)}^2 
\equiv 
\frac{\mu_V}{6} \sqrt{\frac{\chi}{T c_V}} \,.
\label{eq:gamma}
\eeqn
The case of the trivial dispersion relation ($\omega \equiv 0$) is considered in details in Section~\ref{sec:DHS:3}. 

Due to the apparent analogy of the full (magnetic/vortex/heat) wave mixing with the simpler case of the magnetic/heat wave mixing, the analysis of the full mixing can be easily done. Following Section~\ref{sec:MH}, we find that the eigensystem of the velocity matrix~\eq{eq:Vij:2} gives us the principal velocities (squared) $v_{\MVH,\pm}^2 \equiv \lambda_\pm$ and the principal vectors~\eq{eq:n:plus:minus} determined by the angle $\theta$:
\beqn
v_{\MVH,\pm}^2 & = & \frac{1}{2} \biggl(v_{\MV}^2 + v_{\CHW}^2 +  2 \gamma v_{\MV} v_{\CHW} \cos\beta \nonumber\\
& & \pm \sqrt{\left(v_{\MV}^2 + v_{\CHW}^2 +  2 \gamma v_{\MV} v_{\CHW} \cos\beta\right)^2 - 4 (1-\gamma^2) v_{\MV}^2 v_{\CHW}^2 \sin^2\beta}\, \biggr),
\qquad
\label{eq:v:pm:full} \\
\theta({\bs B},{\bs \Omega}) & = & \arctan \left[\left(1 - \frac{(1-\gamma^2) v^2_{\CHW}({\Omega}) }{v^2_{\MVH,+} ({\bs B},{\bs \Omega})} \right) 
\frac{v_\MV({\bs B},{\bs \Omega}) \sin \varphi({\bs B},{\bs \Omega}) }{\gamma v_{\CHW}({\Omega}) + v_\MV({\bs B},{\bs \Omega}) \cos \varphi({\bs B},{\bs \Omega}) }
\right]\!, \qquad
\label{eq:theta:MVH}
\eeqn
where $\beta$ is an angle between the effective magnetic field ${\bs B}^\eff$, Eq.~\eq{eq:B:eff}, and the angular velocity ${\bs \Omega}$, and the factor $\gamma$ is given in Eq.~\eq{eq:gamma}. The mixed waves propagate with velocities $v_{\MVH,\pm}$ along the principal vectors~\eq{eq:n:plus:minus} and \eq{eq:theta:MVH}. Here the subscript ``MVH'' stands for the mixing of the Chiral {\underline M}agnetic--{\underline V}ortical--{\underline H}eat Wave. 

All equations of the end of Section~\ref{sec:MH} can now be applied to the full-wave mixing by making the substitution ${\bs B} \to {\bs B}^{B\Omega}$ where the effective field ${\bs B}^{B\Omega}$ is given in Eq.~\eq{eq:B:eff}. In particular, we conclude that  the direction of the wave vector ${\bs k}$ of the mixed wave and the direction of its velocity ${\bs v}$ do not generally coincide with each other.

If the vectors ${\bs B}$ and ${\bs \Omega}$ are collinear, then the mixed wave propagates along these vectors with (the absolute value of) the velocity 
\beqn
v^{\|,\pm}_{\MVH} & = & \left\{(v_\CVW \pm v_\CMW)^2 
+ \left[1 + {\left(\frac{\mu_V}{\pi T}\right)}^2  \Bigl(1 \pm  \frac{v_\CMW}{v_\CVW}\Bigr)\right] v^2_\CHW\right\}^{\frac{1}{2}}\,,
\label{eq:v:pm:parallel}
\label{eq:v:MVH}
\eeqn
where upper (lower) sign corresponds to parallel (antiparallel) orientation of the vectors $e {\bs B}$ and $\mu_V {\bs \Omega}$, and pure magnetic, vortical and heat velocities are given in Eqs.~\eq{eq:v:CMW}, \eq{eq:v:CVW} and \eq{eq:v:CHW}, respectively.

Notice that if the magnetic field takes a very specific value~\eq{eq:B:special} then the Chiral Magnetic and Chiral Vortical Waves disappear completely and the Chiral Heat Wave remains the only gapless collective mode in the system. Indeed, at this strength~\eq{eq:B:special} the effective magnetic field ${\bs B}^\eff$ vanishes~\eq{eq:B:eff}, and the matrix ${\hat {\bs M}}$ -- which determines the propagations of fluctuations~\eq{eq:hat:M:3} -- gets drastically simplified as it has now only two elements corresponding to the Chiral Heat Wave~\eq{eq:mixing:EA:2}. Basically, for this value of the magnetic field~\eq{eq:B:special}, the magnetic and vortical waves exactly cancel each other implying $v_\CVW = \mp v_\CMW$ and, consequently, $v^{\|,\pm}_{\MVH} \equiv v_\CHW$ in Eq.~\eq{eq:v:pm:parallel}. The remaining heat wave propagates with the standard heat velocity~\eq{eq:v:CHW}, which is affected neither by the presence of the magnetic field ${\bs B}$ nor by the rotation ${\bs \Omega}$.

At the end of this section let us consider briefly the behavior the of mixed magnetic-vortical-heat wave in two specific cases.

The fate of the mixed magnetic-vortical-heat wave in the low-temperature limit can easily be seen from the structure of matrix~\eq{eq:hat:M:3} which describes the generation of the anomalous currents by charge (energy) density fluctuations. At low temperature the heat wave disappears while the magnetic and vortical waves still exist and remain coupled to each other. The energy current is replaced by the mass current along the direction of the effective magnetic field~\eq{eq:B:eff}. The associated mass wave does not, however, influence the velocity of the coupled magnetic-vortical wave because in this limit the magnetic-vortical-heat velocity~\eq{eq:v:pm:full} reduces to the magnetic-vortical velocity~\eq{eq:v:MV}. 

If one keeps temperature nonzero but stops rotation (${\bs \Omega} = 0$) then both the vortical and heat waves formally disappear leaving the CMW alone with the matter wave. According to the form of the mixing matrix~\eq{eq:hat:M:3}, the mass wave is generated by the axial component $\delta \rho_A$ of the CMW.  The mass wave influences neither the velocity nor the chiral content of the CMW neither at zero temperature nor at finite temperature.

\section{Non-propagating diffusive modes: Dense Hot Spots}
\label{sec:DHS}

So far we discussed sound-like collective modes which correspond to coherent propagation of the vector charge density, axial charge density and/or energy density waves along the axis of magnetic field and/or angular velocity vector (or the combinations of the latter two). In the lowest, linear order in momentum these waves possess the linear dispersion relations, $\omega = \pm v k_z$, with the corresponding velocities~$v$. However, in certain environments we have also found the presence of zero-frequency solutions, $\omega = 0$. In this section we discuss these solutions and demonstrate that they describe certain non-propagating diffusive configurations of energy and vector charge densities. These solutions exist due to interplay between axial and mixed gauge-gravitational anomalies in the system.

\subsection{Rotating hot dense fluid in the absence of magnetic field}
\label{sec:DHS:1}

In Sect.~\ref{sec:VH} we have demonstrated that the vortical wave mixes with the heat wave in a rotating finite-density fluid at finite temperature in the absence of magnetic field. In addition, we have observed a new branch of solutions corresponding to the identically zero dispersion~\eq{eq:omega:0} in the linear order of the wave vector~$\bs k$. This new mode is obviously a non-propagating object since its velocity is identically zero in the reference frame defined by the chemical potential ${ \mu}_V$:
\beqn
{\bs v}_{\DHS}({\bs k}) = \frac{\partial \omega ({\bs k})}{\partial {\bs k}} \equiv 0\,.
\eeqn
We call this diffusive mode ``the Dense Hot Spot'' (DHS) because this mode has an excess both in the vector charge density and in the thermal energy density. According to Eqs.~\eq{eq:hat:M} and \eq{eq:diff:1} the density fluctuations in DHS are related to each other:
\beqn
\delta \epsilon(z)  = - \frac{3 c_V { \mu}_V}{\pi^2 { T} \chi}  \, \delta  \rho_V(z)\,, \qquad\ \delta \rho_A(z) = 0\,.
\label{eq:DHS:epsilon:qV}
\eeqn
The fluctuations in energy and the vector charge densities have mutually opposite signs while the fluctuation in the axial charge density is identically zero in the DHS. 

The mentioned example of the DHS corresponds to fluctuations in vector and energy densities~\eq{eq:DHS:epsilon:qV} which (in the linear order in $\bs k$) do not generate fluctuations of currents: 
\beqn
\delta {\bs j}_V = \delta {\bs j}_A = \delta {\bs j}_E = 0\,. 
\label{eq:currents:zero}
\eeqn
Therefore, the DHS is not a propagating diffusive mode. We would like to stress that the DHS is not a standing wave which could be composed from two counter-propagating pure vortex and heat waves. 

Anticipating the inevitable dissipation of the vector charge and thermal diffusion, we expect that the dispersion relation of the Dense Hot Spots should contain, to the lowest order in momenta, the dissipative term only:
\beqn
\omega_{\DHS}(k_z) = - i D_\| k^2_z  + \dots\,.
\eeqn
Once this fluctuation is created it would diffuse without propagation. The shape of the DHS is not determined by the linear anomalous relations, so that for small densities it can be an arbitrary function of the longitudinal coordinate $z$.

Thus we arrive to the following convenient definition for the Dense Hot Spot:
\begin{itemize}
\item[(i)] The DHS consists of non-propagating diffusive lumps in the energy and vector charge densities which are related to each other;
\item[(iii)] The axial charge density in the DHS is identically zero;
\item[(ii)] The DHS does not generate vector, axial and thermal energy currents via anomalous transport laws~\eq{eq:currents:zero}. 
\end{itemize}

\subsection{Rotating hot zero-density fluid in magnetic field}
\label{sec:DHS:2}

Now let us consider the case of rotating (${\bs \Omega} \neq 0$) finite-temperature ($T\neq 0$) neutral (${ \mu}_V = 0$) fluid subjected to an external magnetic field (${\bs B} \neq 0$). This setup is relevant to the quark-gluon plasma created in noncentral heavy-ion collisions as the plasma is set into rotation due to noncentrality of the collision, while the charged nature of the colliding ions exposes the plasma to the external magnetic field. Geometrically, both the axis magnetic field ${\bs B}$ and the the angular velocity vector ${\bs \Omega}$ are co-aligned in noncentral collisions. However, for the sake of generality, we will first consider below the case of arbitrary orientation of the vectors ${\bs B}$ and ${\bs \Omega}$.

As we have already discussed in Sect.~\ref{sec:MH}, in this environment  ($T\neq 0$, ${ \mu}_V = 0$, ${\bs B} \neq 0$, ${\bs \Omega} \neq 0$) both chiral magnetic and heat waves emerge and they mix with each other. In addition to the coupled chiral magnetic-heat wave we have also found a branch of the spectrum with the zero dispersion relation~\eq{eq:omega:0:MH}. This is yet another example of a non-propagating diffusive mode, the Dense Hot Spot. 

The charge and energy content of the DHS can be found from Eqs.~\eq{eq:mixing1:jV:2}, \eq{eq:mixing2:jA:2} and \eq{eq:jE:2}. Since the spot should generate no anomalous currents, we find from Eqs.~\eq{eq:mixing1:jV:2} and \eq{eq:jE:2} that the axial density in the spot is zero, $\delta \rho_A = 0$. Equation~\eq{eq:mixing2:jA:2} implies the following relation between charge and energy fluctuations valid in the linear order of momentum $\bs k$:
\beqn
\frac{e ({\bs B} \cdot {\bs k})}{2 \pi^2 \chi} \delta  \rho_{V,{{\bs k}}}({\bs x}) + \frac{{ T} ({\bs \Omega} \cdot {\bs k})}{3 c_V} \delta \epsilon_{{\bs k}}({\bs x}) = 0\,.
\label{eq:DHS:vector}
\eeqn

\begin{figure*}[!thb]
\begin{center}
\includegraphics[scale=0.31,clip=true]{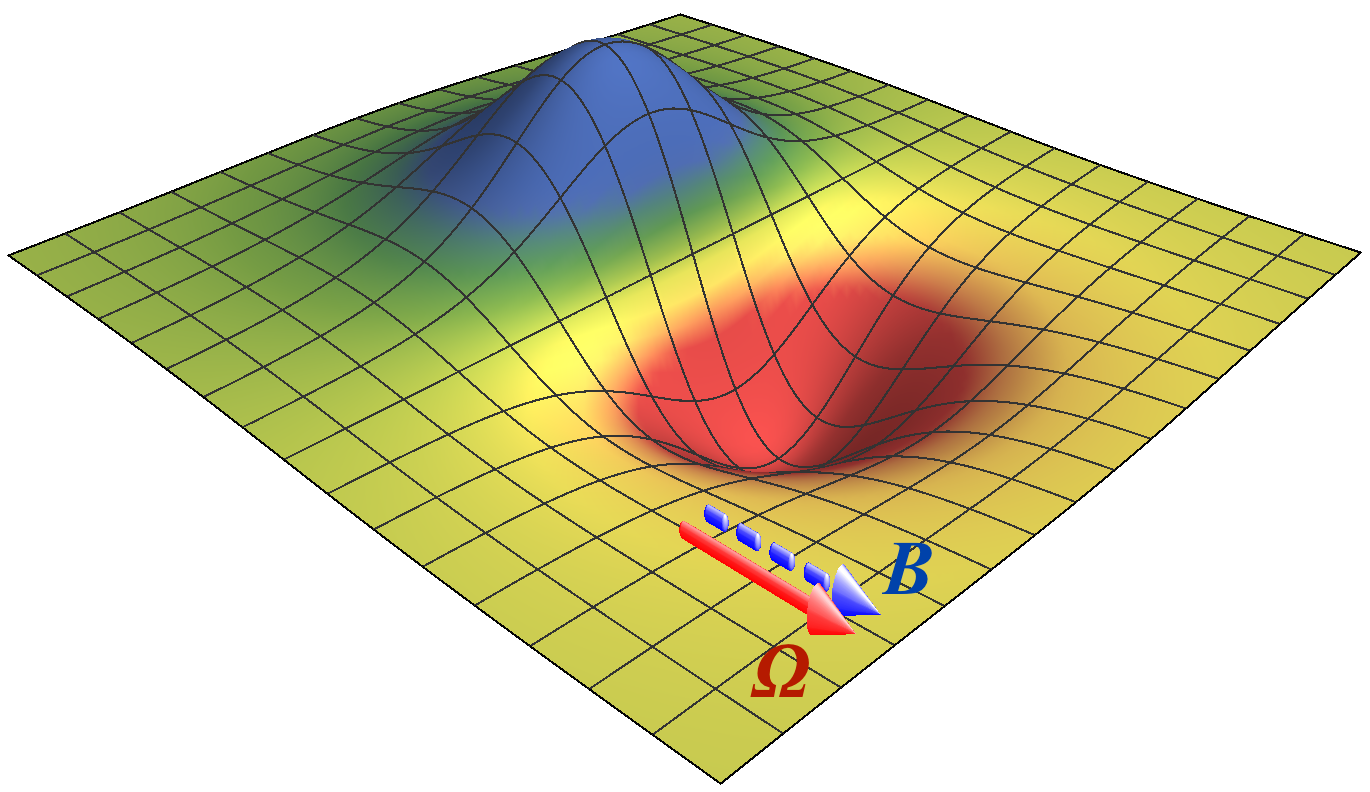} 
\hskip 1mm
\includegraphics[scale=0.31,clip=true]{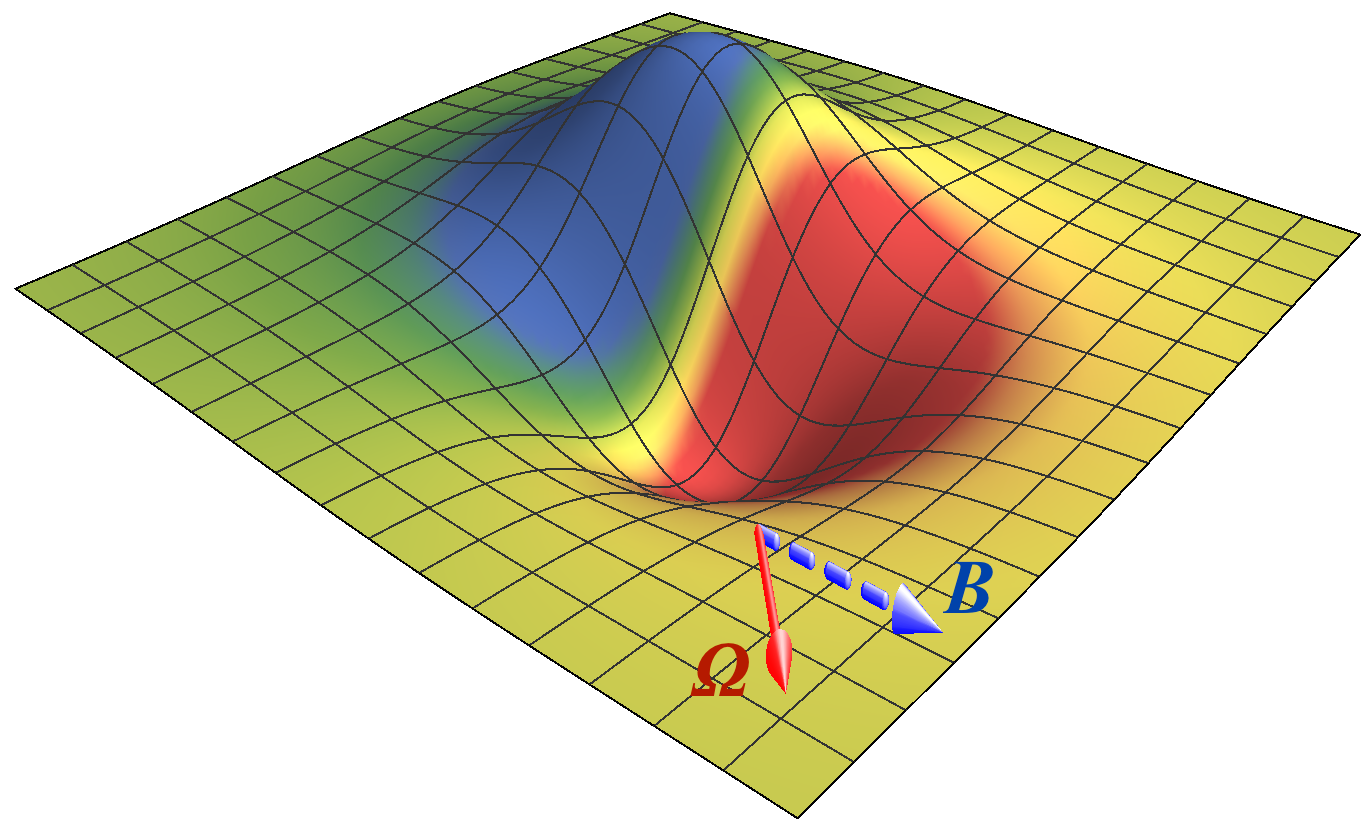}
\hskip 1mm
\includegraphics[scale=0.31,clip=true]{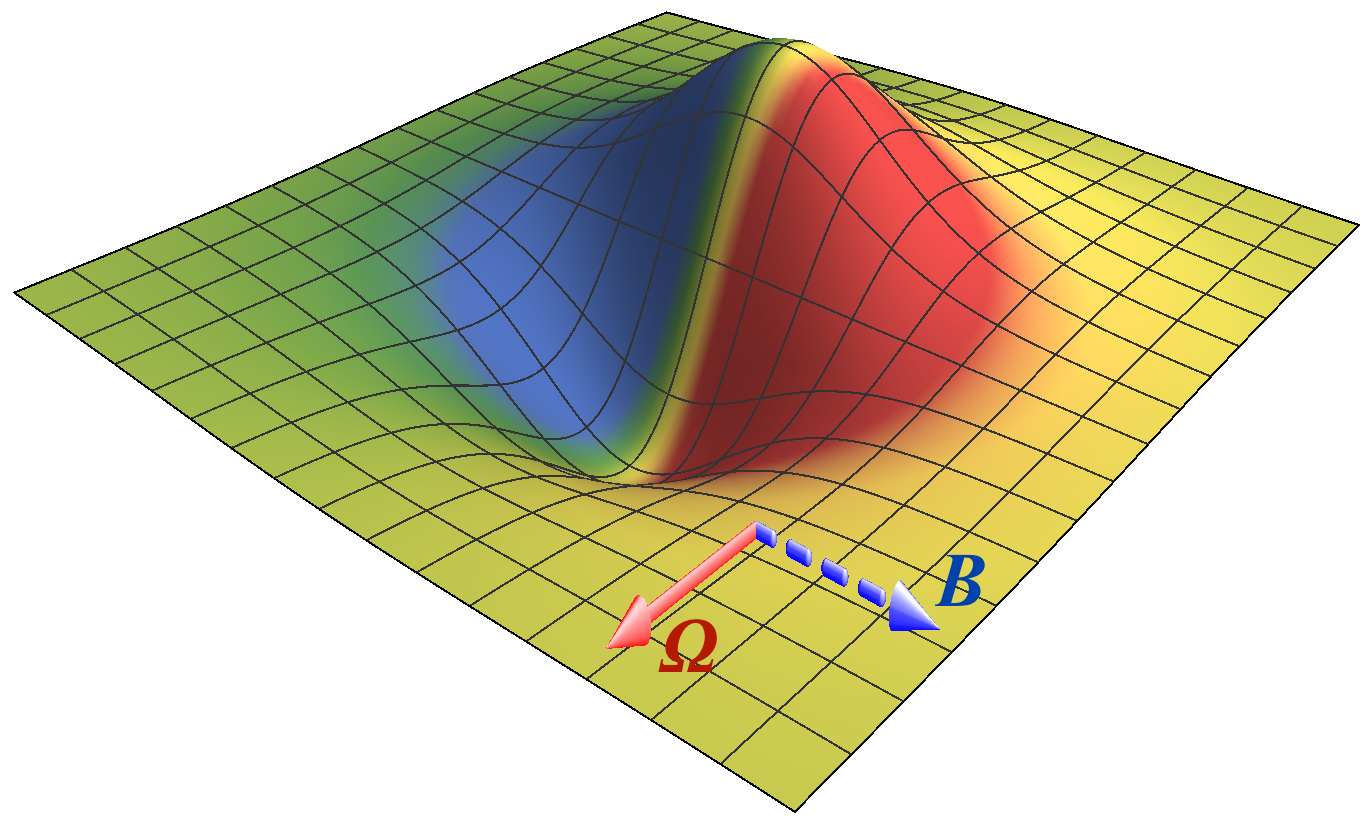}
\hskip 1mm
\includegraphics[scale=0.31,clip=true]{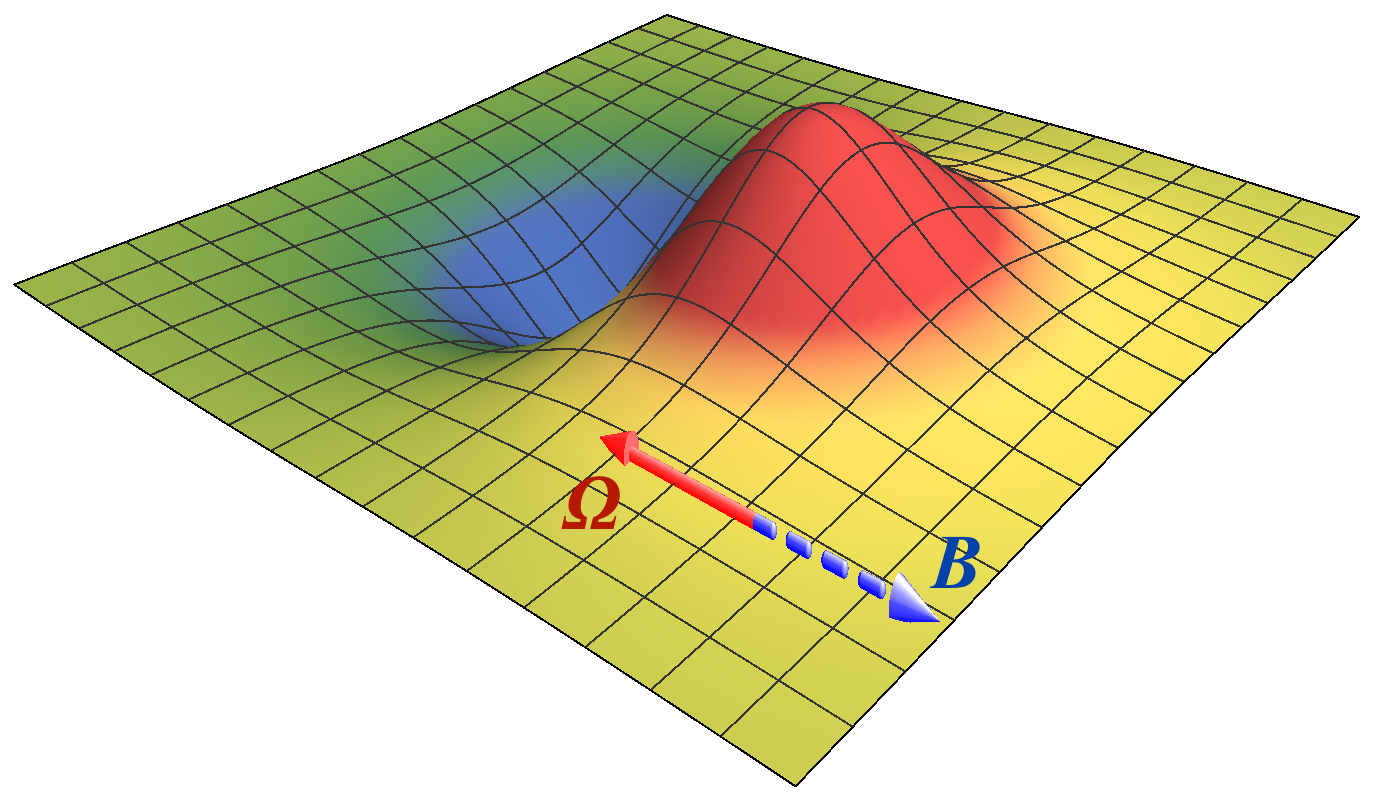}
\end{center}
\caption{Illustration of the Dense Hot Spots in rotating hot fluid in a background magnetic field. Each plot shows the vector density fluctuation $\delta \rho_V$ ($z$-axis magnitude) and the thermal energy fluctuation $\delta \epsilon$ (shown by colors from red/hot to blue/cold)  in coordinate space. The axial density is identically zero, $\delta \rho_A \equiv 0$. The blue dashed (red solid) arrow points to the direction of the magnetic field ${\bs B}$ (the angular velocity ${\bs \Omega}$).}
\label{fig:DHS}
\end{figure*}

The relation~\eq{eq:DHS:vector} between the energy density $\epsilon_{{\bs k}}$ and the vector charge density $\rho_{V,{{\bs k}}}$ depends on the mutual orientation of the magnetic field ${\bs B}$, the angular velocity ${\bs \Omega}$ and the wave vector ${\bs k}$ of the DHS. In Eq.~\eq{eq:DHS:vector} the subscript ${\bs k}$ indicates that the energy density fluctuation $\delta \epsilon_{{\bs k}}({\bs x}) = C_{\bs k} \cos\bigl( ({\bs k} \cdot {\bs x}) + \alpha_{k} \bigr)$, and similarly  the vector density $\delta  \rho_{V,{{\bs k}}}$, is defined for certain wave vector ${\bs k}$. Illustrations of the diffusive Dense Hot Spots for a set of mutual orientations of the angular velocity and the magnetic field are shown in Fig.~\ref{fig:DHS}.

\subsection{Rotating hot dense fluid in magnetic field}
\label{sec:DHS:3}

Finally, let us consider the most general situation, when the density, temperature, magnetic field and angular velocity are all nonzero. The analysis of the DHS can be done similarly to the analysis of the previous section. According to Eqs.~\eq{eq:mixing1:jV:3}, \eq{eq:mixing2:jA:3} and \eq{eq:jE:3} one arrives to a density-energy constraint similar to Eq.~\eq{eq:DHS:vector}, in which the magnetic field ${\bs B}$ is replaced by the effective magnetic field~\eq{eq:B:eff}: $e {\bs B}^\eff \equiv e {\bs B} + 2 { \mu}_V {\bs \Omega}$. Thus, we come to a conclusion that the vector density fluctuations and the energy density fluctuations in the DHSs of rotating hot fluids in the background of magnetic field are qualitatively the same (up to the redefinition of the magnetic field, ${\bs B} \to {\bs B}^\eff$) for the cases of zero ($\mu_V = 0$) and nonzero  ($\mu_V \neq 0$) background densities. 

In addition of the DHS-like fluctuations, the rotating dense system of chiral fermions may host a ``classical'' non-propagating DHS mode, in which the magnitudes of vector charge and energy densities are not limited to small values. Such ``giant'' spots are realized in a special case, when the magnetic field and angular velocities are collinear to each other: ${\bs B} \| {\bs \Omega}$. At zero axial chemical potential, $\mu_A = 0$, the anomalous vector current~\eq{eq:CME:jV}, \eq{eq:sigma:VB:AB}, \eq{eq:CVE:jVA}, \eq{eq:sigma:CVE} and the anomalous energy current~\eq{eq:jE}, \eq{eq:sigma:EB}, \eq{eq:sigma:EV} are both zero so that the non-propagation condition for these quantities is satisfied automatically. As for the anomalous axial current, it is given by Eqs.~\eq{eq:CSE:jA}, \eq{eq:sigma:VB:AB}, \eq{eq:CVE:jVA} and \eq{eq:sigma:CVE}:
\beqn
{\bs j}_A =
\frac{\mu_V}{2 \pi^2} e {\bs B} + \left(\frac{T^2}{6} + \frac{\mu_V^2}{2 \pi^2} \right) {\bs \Omega}\,.
\label{eq:mixing2:jA:2:app3}
\eeqn
Taking ${\bs B} = B {\bs e}_z$ and ${\bs \Omega} = \Omega {\bs e}_z$ along the same axis, one finds that the axial current~\eq{eq:mixing2:jA:2:app3} vanishes in the DHS if the local temperature and local value of the chemical potential satisfy the following relation:
\beqn
3 \mu_V(x) eB + 3 \mu_V^2(x) \Omega + \pi^2 T^2(x) \Omega = 0\,.
\label{eq:DHS}
\eeqn
Similarly to equations for the chiral waves, this relation for the chiral diffusive spot is applicable only in the long-wavelength limit. 

Notice that according to Eq.~\eq{eq:DHS} the ``giant DHS'' at zero temperature is realized at the particular value of the chemical potential: 
\beqn
\mu_V = - \frac{eB}{\Omega}\,.
\eeqn

Summarizing this section we conclude that the Dense Hot Spots are static diffusive modes which appear due to interplay of the axial and gauge-gravitational anomalies in the environment that couples the heat wave either to the vortex wave (Section~\ref{sec:DHS:1}), or to the magnetic wave (Section~\ref{sec:DHS:2}) or to the common magnetic-vortex wave (Section~\ref{sec:DHS:3}). The DHS may only exist in the rotating chiral medium subjected to an external magnetic field.

\section{Summary and conclusions}
\label{sec:summary}

We have demonstrated that a hot rotating fluid/plasma of chiral fermions possesses a new gapless collective excitation, the Chiral Heat Wave, which is associated with the coherent propagation of thermal energy density and chiral charge density waves along the axis of rotation. The heat wave propagation is a cyclic process: the mixed gauge-gravitational anomaly converts a perturbation in the axial charge density into energy current directed along the axis of rotation. Then the energy current heats the chiral medium which generates an excess of the axial current along the same axis. Finally, the axial current leads again to an excess in the axial charge density further along the rotation axis and the processes repeats again. 

At finite density a rotating fluid supports also the Chiral Vortical Wave~\cite{ref:CVW} which mixes with the Chiral Heat Wave at nonzero temperature. Moreover, in the presence of an external magnetic field the system may also host the Chiral Magnetic Wave~\cite{ref:CMW} which, in a rotating fluid, should couple to the heat wave and which may also mix with the vortical wave. Since the mentioned vortical and magnetic waves are propagating due to similar cyclic conversion of the vector and axial densities, the mixed vortical-heat and magnetic-heat waves involve fluctuations of all three (vector, axial and energy) currents and their densities, which appear in different proportions depending on external conditions (temperature, density, rotation and magnetic field).
\begin{table}[!htb]
\begin{center}
\begin{tabular}{c|c|c|c|c|}
\cline{2-5} & \multicolumn{4}{c|}{$T \neq 0$ } \\
\cline{2-5} & \multicolumn{2}{c}{${\bs \Omega} = 0$ } & \multicolumn{2}{|c|}{${\bs \Omega} \neq 0$ } \\
\cline{2-5} & ${\bs B} = 0$ &  ${\bs B} \neq 0$        & ${\bs B} = 0$ &  ${\bs B} \neq 0$ \\
\hline
\multicolumn{1}{|c|}{$\mu_V = 0$}       &   --     &    M \eq{eq:v:CMW}    &    H \eq{eq:v:CHW}   &    M+H \eq{eq:v:pm}, ${\cal A}$, DHS \\
\hline
\multicolumn{1}{|c|}{$\mu_V \neq 0$}  &   --     &    M+{\large{$\mu$}} \eq{eq:v:CMW}&     V+H \eq{eq:v:VH}, DHS   &   M+V+H \eq{eq:v:pm:full}, ${\cal A}$, DHS \\
\hline
\multicolumn{5}{c}{} \\[0mm]
\cline{2-5} & \multicolumn{4}{c|}{$T = 0$ } \\
\cline{2-5} & \multicolumn{2}{c}{${\bs \Omega} = 0$ } & \multicolumn{2}{|c|}{${\bs \Omega} \neq 0$ }  \\
\cline{2-5} & ${\bs B} = 0$ &  ${\bs B} \neq 0$        & ${\bs B} = 0$ &  ${\bs B} \neq 0$ \\
\hline
\multicolumn{1}{|c|}{$\mu_V = 0$}    &   --     &    M   \eq{eq:v:CMW}   &    --       &    M    \eq{eq:v:CMW}     \\ 
\hline
\multicolumn{1}{|c|}{$\mu_V \neq 0$}  &   --     &    M+{\large{$\mu$}}   \eq{eq:v:CMW}     &     V+{\large{$\mu$}}    \eq{eq:v:CVW}   &   M+V+{\large{$\mu$}}     \eq{eq:v:MV}, DHS \\
\hline
\end{tabular}
\end{center}
\caption{Physical conditions required for existence of the Chiral Magnetic (M), Vortical (V), Heat (H) and Mass ($\mu$) Waves, various mixings of these waves and the Dense Hot Spots (DHS) at finite temperature $T \neq 0$ (the upper table) and zero temperature $T=0$ (the lower table). The numbers in round brackets point out to expressions for the corresponding velocities. The pure Chiral Magnetic Wave and Chiral Heat Wave exist at zero density only. Strictly speaking, the Chiral Vortical Wave does not exist alone as it is always coupled either to the Chiral Heat Wave at finite temperature or to the mass wave at zero temperature. The admixture of the mass ($\mu$) wave to any other (M, V, H) wave does not influence the velocity of the mixed wave propagation. The label ${\cal A}$ indicates that the Chiral Magnetic-Heat Wave propagates anisotropically as its phase and group velocities are pointing out, generally, to different directions.}
\label{tbl:T}
\end{table}

In it important to stress that the mixed waves have, in general, different velocities compared to the velocities of the individual constituent waves. For example, if the magnetic field ${\bs B}$ and the angular velocity ${\bs \Omega}$ are pointing to the same direction (and assuming for simplicity that the electric charge $e$ and vector chemical potential $\mu_V$ are both non-negative), then the velocities of the mixed magnetic-vortical~\eq{eq:v:MV}, magnetic-heat~\eq{eq:v:MH}, vortical-heat~\eq{eq:v:VH} and magnetic-vortical-heat~\eq{eq:v:MVH} waves are, respectively, as follows:
\beqn
v^\|_{\MV}   & = & v_\CMW + v_\CVW\,, \nonumber\\
v^\|_{\MH}  & = & \sqrt{v_{\CMW}^2 + v_{\CHW}^2}\,, \nonumber\\
v^\|_{\VH}  & = & \sqrt{v_\CVW^2 + \left[1 + 3 {\left(\frac{\mu_V}{\pi T}\right)}^2 \right]v_\CHW^2}\,, 
\label{eq:velocities:mix}\\
v^\|_{\MVH} & = & \left\{\left[1 + \left( \frac{\mu_V}{\pi T} \right)^2 \left(1+ \frac{v_\CMW}{v_\CVW} \right) \right]v_\CHW^2 
+ \left(v_\CMW + v_\CVW \right)^2 \right\}^{\frac{1}{2}}\,, \nonumber
\eeqn
where 
\beqn
\begin{split}
v_{\CMW} = \frac{eB}{2 \pi^2 \chi}\,, \qquad
v_{\CVW} = \frac{{ \mu}_V \Omega}{\pi^2 \chi}\,, \qquad
v_{\CHW} = \sqrt{\frac{{ T}^3}{c_V \chi}} \frac{\Omega}{3}\,,
\end{split}
\label{eq:summary:V:pure}
\eeqn
are the velocities of the ``pure'' Chiral Magnetic~\eq{eq:v:CMW}, Chiral Vortical~\eq{eq:v:CVW} and Chiral Heat~\eq{eq:v:CHW} Waves, respectively. In Eq.~\eq{eq:velocities:mix} the superscript ``$\|$'' indicates that the velocities are shown for the special case when the magnetic field and the angular velocity are parallel to each other, ${\bs B} \| {\bs \Omega}$. We consider slowly rotating system at small chemical potential ($\Omega \ll T, \mu_V \ll T$), so that $O(\Omega^2)$ and $O(\mu^2_V)$ terms in energy density~\eq{eq:epsilon} are neglected.

Equations~\eq{eq:velocities:mix} indicate that the mixing of the Chiral Magnetic Wave with the Chiral Heat Wave makes the velocity of the original CMW higher. In other words, the magnetic wave (which appears as a result of the axial anomaly) propagates faster at finite temperature due to the presence of the mixed gauge-gravitational anomaly. The same is true for the mix of the Chiral Vortical Wave and the Heat Wave: the coupling to the energy density wave makes the vortical wave faster. Notice that, strictly speaking, the Chiral Vortical Wave never exists alone: it is always coupled either to the Chiral Heat Wave (at finite temperature) or to the mass wave (at zero temperature). Due to this inevitable coupling the velocity of the Chiral Vortical Wave at finite temperature~\eq{eq:v:VH} is always higher compared to the zero-temperature expression or to idealized formula for a pure Chiral Vortical Wave given by the second formula in Eq.~\eq{eq:summary:V:pure}]. 

As for the mix of the magnetic and vortical waves, the result of the coupling between these waves depends on the relative signs of the magnetic field and the angular velocity. The mixed wave may propagate faster, slower or even stop propagating at all. The latter happens if the the magnetic field and the angular frequency obey Eq.~\eq{eq:B:special}. 
There are also other effects of the wave mixing. If the angular velocity of the chiral fluid is not collinear to the axis of magnetic field, then there exists a mixed Heat-Magnetic wave which propagates anisotropically: its phase and group velocities are, in general, not parallel to each other (in other words, the wave vector ${\bs k}$ and the velocity ${\bs v}$ of the mixed wave are not collinear as it is illustrated in Fig.~\ref{fig:waves} of Section~\ref{sec:MH}). The same is true for the triple, Heat-Magnetic-Vortical wave mixing discussed in Section~\ref{sec:mix:MVH}. As for the Heat-Vortical wave (Section~\ref{sec:VH}) and the Magnetic-Vortical wave  (Section~\ref{sec:MV}), their phase and group velocities are always parallel to each other.

A mixing of the heat wave either with the magnetic wave or with the vortical wave or with both these waves leads also to appearance of the diffusive modes, the Dense Hot Spots which are {\sl non-propagating} thermal fluctuations with zero chiral charge density but with nonzero vector charge density. In the first, linear order in momentum, the DHSs possess identically zero dispersion law ($\omega = 0$) so that the corresponding phase and group velocities are identically zero. The fluctuations in energy density and in vector charge density of a DHS are related to each other in such a way that they generate no anomalous vector, axial and energy currents. 

In Table~\ref{tbl:T} we briefly summarize the physical conditions at which the pure Chiral Magnetic/Vortical/Heat Waves, the corresponding inter-wave mixings and the diffusive Dense Hot Spots may appear. 

We expect that our results should be relevant to noncentral heavy-ion collisions which create rotating fireballs of hot quark-gluon plasma subjected to a strong magnetic field. In this environment all discussed sound-like modes, namely the Chiral Magnetic, Chiral Vortical and Chiral Heat Waves should exist. We expect that these waves should inevitably mix with each other and form a single collective wave which could be either Chiral Magnetic-Heat or Vortical-Heat or Magnetic-Vortical-Heat Wave, depending on the baryon density, the strength of magnetic field and the angular velocity of the fireball ({\it cf.} Table~\ref{tbl:T}). The common collective wave propagates as a coherent excitation in vector, axial and thermal energy densities. Since in a typical noncentral heavy-ion collision the axis of the magnetic field is co-aligned with the angular velocity, the group and phase velocities of the common collective wave should coincide. 

\acknowledgments

This work was partially supported by Far Eastern Federal University grant 13-09-0617-m\_a and by the Sino-French Cai Yuanpei Exchange Program (Partenariat Hubert  Curien). Useful communications with Xu-Guang Huang, Karl Landsteiner, Jinfeng Liao and Igor Shovkovy are gratefully acknowledged. The author thanks the Theory Division of the Institute of High Energy Physics (Beijing, China) for the kind hospitality extended to him during his stay.

{\sl Note added}: --- After this paper was submitted, a new preprint has appeared~\cite{Frenklakh:2015fzc} where the existence of the coupled Chiral Magnetic-Vortical Wave, predicted in Section~\ref{sec:MV} of our paper, was confirmed in a kinetic theory. In particular, our Eq.~\eq{eq:v:MV} for the velocity of the Magnetic-Vortical wave was re-derived.

\end{document}